\documentclass[%
reprint,
superscriptaddress,
amsmath,amssymb,
aps,
floatfix,
longbibliography
]{revtex4-2}

\usepackage{graphicx}
\usepackage[breaklinks]{hyperref}

\usepackage{times}
\usepackage{dcolumn}
\usepackage{float}
\usepackage[dvipsnames]{xcolor}

\hypersetup{colorlinks=true,linkcolor=blue,citecolor=blue,urlcolor=blue}

\usepackage{mathtools}
\usepackage[normalem]{ulem}

\usepackage[whole]{bxcjkjatype}

\DeclarePairedDelimiter{\tripleangle}
  {\langle\!\langle\!\langle}
  {\rangle\!\rangle\!\rangle}

\makeatletter
\def\maketitle{
\@author@finish
\title@column\titleblock@produce
\suppressfloats[t]}
\makeatother

\begin{document}


\title{Phase diagram of the Kitaev-Heisenberg-$\Gamma$ model: \\ Classical and quantum magnetism, frustration, and subdominant interactions}

\author{Kiyu Fukui}
\email{k.fukui@aion.t.u-tokyo.ac.jp}
\affiliation{Department of Physical Sciences, Ritsumeikan University, Kusatsu, Shiga 525-8577, Japan}

\author{Yukitoshi Motome}
\affiliation{Department of Applied Physics, The University of Tokyo, Bunkyo, Tokyo 113-8656, Japan}

\date{\today}

\begin{abstract}
The Kitaev spin liquid provides a rare example of exactly solvable quantum spin liquid states. Intensive research over the past two decades has identified a variety of its candidate materials. In real materials, however, the Kitaev interaction is inevitably accompanied by additional magnetic interactions such as the Heisenberg and $\Gamma$ interactions. These interactions often induce magnetic ordering at low temperatures, making it essential to clarify their effects in the search for and design of Kitaev spin liquid candidate materials. In this study, we revisit the ground-state phase diagram of the Kitaev-Heisenberg-$\Gamma$ model from both classical and quantum perspectives, using state-of-the-art numerical techniques. In the classical case, we reveal a \textit{zoo of noncollinear orders}, where a variety of noncollinear multiple-$Q$ magnetic orders with and without incommensurate modulations emerge. In the quantum case, we unravel that quantum fluctuations suppress many of the competing orders found in the classical case, resulting in a reduced number of dominant incommensurate orders. We further identify \textit{highly frustrated} regions, where spiral spin liquid states as well as new magnetically ordered states are potentially stabilized by other additional magnetic interactions. Our results provide a comprehensive perspective on the Kitaev-Heisenberg-$\Gamma$ model for both classical and quantum spins and offer a valuable guide not only for interpreting experimental results on candidate materials, but also for searching and designing new materials to realize the Kitaev spin liquid.
\end{abstract}

\maketitle

\section{Introduction}\label{sec:intro}

Quantum spin liquids (QSLs) are exotic states of matter in which strong frustration and quantum fluctuations suppress long-ranged magnetic ordering down to zero temperature~\cite{Balents2010, Lacroix, Savary2017, Zhou2017, Diep}. Unlike conventional magnetically ordered states characterized by spontaneous symmetry breaking and local order parameters, QSLs harbor long-range quantum entanglement, fractionalized excitations, and emergent gauge structures. These remarkable properties make QSLs not only a profound challenge in condensed matter physics but also a promising resource for quantum computation and quantum information science~\cite{Kitaev2003, Kitaev2006, Nayak2008, Pachos, Zeng}.
 
Despite more than half a century of research since P. W. Anderson's proposal~\cite{Anderson1973, Fazekas1974}, well-established examples of QSLs remain limited, and the realization of the QSLs in most of the candidate models and materials is still under debate.
Among various theoretical proposals for QSLs, the Kitaev model defined on the honeycomb lattice occupies a unique position, as it provides a rare example of exactly solvable QSL ground states in more than one dimension~\cite{Kitaev2006}. In this model, bond-dependent Ising-type interactions between spin-$1/2$ moments generate strong frustration and fractionalize spins into itinerant Majorana fermions coupled to a $\mathbb{Z}_2$ gauge field.
Stimulated by the seminal proposal by Jackeli and Khaliullin that the Kitaev-type bond-dependent anisotropic interactions can be realized in spin-orbit coupled Mott insulators via edge-sharing octahedra geometry~\cite{Jackeli2009}, intensive experimental and theoretical
efforts have been devoted to exploring Kitaev candidate materials~\cite{Rau2016, Winter2017, Takagi2019, Motome2020a, Trebst2022, Kim2022, Rousochatzakis2024, Garlea2025, Matsuda2025, Aristizabal2026}.
Early candidates focused on $5d$ iridium compounds with a honeycomb structure, with Na$_2$IrO$_3$~\cite{Chaloupka2010, Singh2010, Singh2012, Comin2012, Chaloupka2013, Foyevtsova2013, Sohn2013, Katukuri2014, Yamaji2014, HwanChun2015, Winter2016} and $\alpha$-Li$_2$IrO$_3$~\cite{Singh2012, Chaloupka2013, Winter2016, Williams2016} investigated as the first realizations.
The $4d$ compound $\alpha$-RuCl$_3$ subsequently emerged as a
particularly prominent candidate~\cite{Plumb2014, Kim2015, Kubota2015, Winter2016, Yadav2016, Banerjee2016, Sinn2016, Kim2022b, Matsuda2025, Moller2025, Aristizabal2026}. A half-integer quantized thermal Hall conductance~\cite{Kasahara2018, Yamashita2020, Yokoi2021, Bruin2022, Kasahara2022, Imamura2024} and field-angle-dependent specific heat~\cite{Tanaka2022, Imamura2025} were reported for this compound, attracting enormous attention as a possible signature of Majorana fermions.

Motivated by progress in these materials, the search has recently been extended to cobalt-based honeycomb materials such as Na$_2$Co$_2$TeO$_6$ and Na$_3$Co$_2$SbO$_6$~\cite{Liu2018, Sano2018, Yan2019, Liu2020, Chen2021, Hong2021, Lin2021, Samarakoon2021, Kim2022, Kim2022a, Li2022, Zhang2022, Halloran2023, Xu2023, Yao2023, Haraguchi2024, Maksimov2025, Bestha2026}, iridium ilmenites~\cite{Haraguchi2018, Haraguchi2020, Jang2021}, $f$-electron compounds~\cite{Jang2019, Jang2020, Xing2020, Ramanathan2021, Daum2021, Ishikawa2022, Jang2024, Okuma2024}, and ruthenium halides in which chlorine in $\alpha$-RuCl$_3$ has been replaced by bromine or iodine~\cite{Salavati2019, Ni2021, Nawa2021, Imai2022, Samanta2023, Zhang2025, Nawa2026, Gretarsson2026}. In addition to $S=1/2$ systems, higher-spin ($S>1/2$) materials have also been proposed. In nickel-based compounds, the extended $S=1$ Kitaev interaction has been discussed theoretically in $A_3$Ni$_2X$O$_6$ ($A =$ Li, Na; $X =$ Bi, Sb)~\cite{Stavropoulos2019}, and inelastic neutron scattering on Na$_2$Ni$_2$TeO$_6$ has revealed signatures consistent with a ferromagnetic Kitaev interaction~\cite{Samarakoon2021}.
Chromium-based materials CrI$_3$ and CrBr$_3$ with $S=3/2$ moments have been discussed as candidates with substantial Kitaev interactions~\cite{Kim2019, Xu2018, Lee2020a, Stavropoulos2021}, and the realization of an antiferromagnetic Kitaev QSL was predicted for strained monolayers of CrSiTe$_3$ and CrGeTe$_3$~\cite{Xu2018, Xu2020}.

\begin{figure*}[t]
    \centering
     \includegraphics[width=\linewidth,clip]{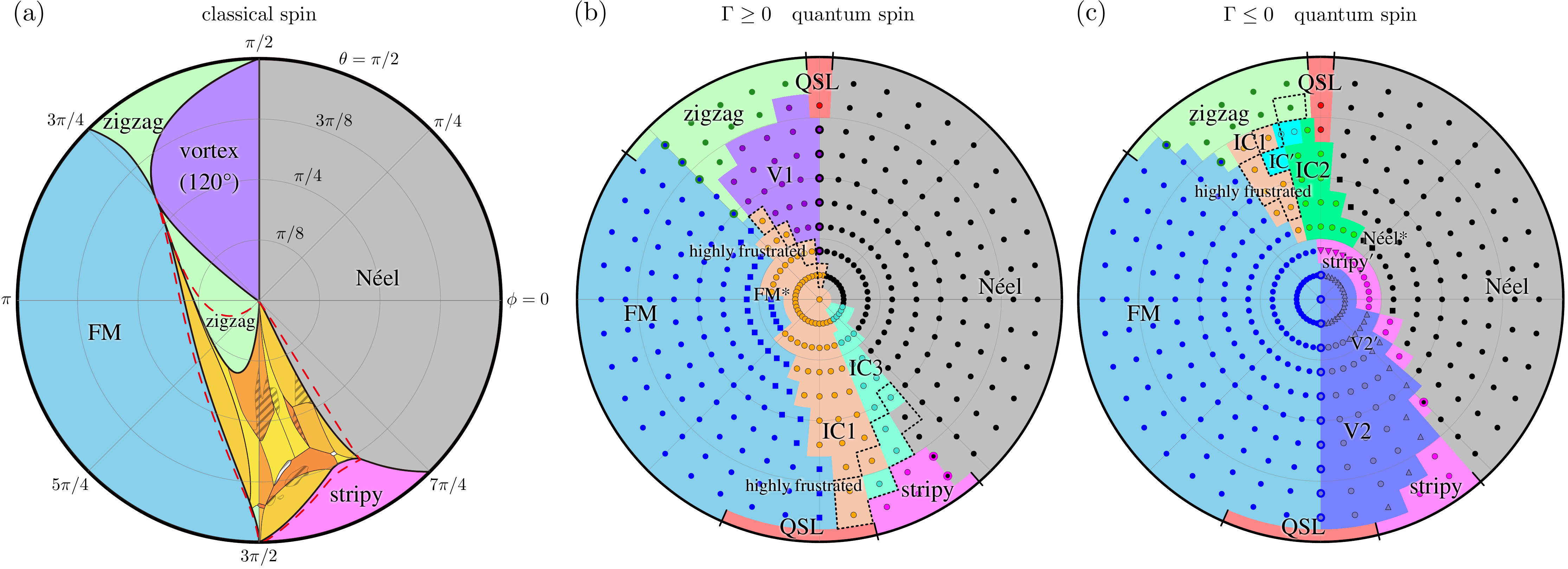}
    \caption{Ground-state phase diagrams of the Kitaev-Heisenberg-$\Gamma$ model in Eq.~\eqref{eq:model}, parametrized by polar angle $\theta$ and azimuthal angle $\phi$ as defined in Eq.~\eqref{eq:coupling}, for (a) classical spins for $\Gamma\geq 0$, (b) quantum spins for $\Gamma\geq 0$, and (c) quantum spins for $\Gamma\leq 0$. In (a), $(J, K, \Gamma)$ can be transformed to $(-J, -K, -\Gamma)$ by a site-dependent time-reversal, and thus only the region with $\Gamma \geq 0$ is shown. Calculations are carried out at $314620$ parameter points. The region enclosed by the red dashed line indicates where the Luttinger-Tisza method loses its validity. The hatched regions indicate where incommensurate magnetic order is suggested.
    In (b) and (c), the black ticks on the outer circles indicate the phase boundaries of the Kitaev-Heisenberg model obtained in the previous study~\cite{Fukui2022a}, with rescaling to match the energy scale of the present calculations. The symbols denote the calculated points, and overlap of symbols representing different phases indicates regions exhibiting phase-boundary--like behavior or suggesting possible phase coexistence. The regions enclosed by the black dashed lines indicate highly frustrated regions (see main text).
    }
    \label{fig:main}
\end{figure*}

Although a large number of candidate Kitaev materials have been proposed, almost all of them exhibit magnetic ordering, typically zigzag or spiral orders, at low temperature due to non-Kitaev interactions.
For realistic materials, such as those described above, perturbation theory and first-principles calculations have shown that magnetic interactions within these materials are not limited to the pure Kitaev interaction: typical examples include the Heisenberg interaction which is isotropic in spin space and the $\Gamma$ interaction which is one of spin off-diagonal interactions.
These interactions induce strong competition between magnetically ordered states and the spin liquid states, as well as among different magnetically ordered states. Therefore, clarifying the effect of these interactions is a central issue in the search for and design of promising materials hosting the Kitaev spin liquid. 

However, since non-Kitaev interactions compromise the solvability of the model, elucidating their effects in detail requires numerical calculations of highly frustrated spin systems, which remains a formidable challenge even today.
Historically, the Kitaev-Heisenberg model was first proposed as a minimal model, and later the importance of the $\Gamma$ interaction was recognized for capturing the magnetism of the candidate materials. Treating the Kitaev, Heisenberg, and $\Gamma$ interactions simultaneously is quite complicated due to the strong frustration arising from their competition. Thus, the ground-state phase diagrams of simpler Kitaev-Heisenberg and Kitaev-$\Gamma$ models have been intensively studied for both classical and quantum spins, leading to substantial progress.
In contrast, the phase diagram of the Kitaev-Heisenberg-$\Gamma$ model --- one of the most representative models for describing Kitaev materials --- remains far from fully understood. In the classical case, previous studies~\cite{Rau2014, Rau2014a} relied on numerical methods which are not able to fully capture the multiple-$Q$ and incommensurate orders expected in the strongly competing region where many candidate materials reside. In the quantum case, available studies have been limited to small clusters~\cite{Rau2014, Rau2014a, Lou2015, Rusnacko2019}, which are insufficient to detect the multiple-$Q$ and incommensurate magnetic orders that require large system sizes, leaving large portions of the phase diagram unsettled. These limitations call for a comprehensive revisitation of the phase diagram using state-of-the-art numerical methods capable of treating large system sizes and detecting incommensurate magnetic orders.

In this paper, we investigate the ground-state phase diagram of the Kitaev-Heisenberg-$\Gamma$ model by performing large-scale numerical calculations for both classical and quantum spin cases. The resulting phase diagrams are summarized in Fig.~\ref{fig:main}.
In the classical case, we revisit the phase diagram using a gradient descent optimization based on an automatic differentiation scheme, which enables unbiased energy minimization~\cite{Pohle2023, Pohle2024, Ikegami2026a}. We reveal that, in the strongly competing region where many candidate materials are located, a variety of noncollinear multiple-$Q$ magnetic orders with and without incommensurate modulations emerge and compete intensely. We comprehensively identify this \textit{zoo of noncollinear orders} through energy derivatives, system size dependence, and scalar spin chirality, substantially revising the previously reported phase diagrams. 
To elucidate the role of quantum fluctuations, we further study the ground-state phase diagram for quantum spins using the pseudo-fermion functional renormalization group (PFFRG) method~\cite{Reuther2010, Buessen2019, Muller2024}. We show that quantum fluctuations suppress many of the competing orders in the classical case, resulting in a reduced number of dominant phases. In particular, we identify regions that may stabilize several incommensurate magnetic orders, and extended
\textit{highly frustrated} regions where the spin susceptibility exhibits ring-like features in momentum space. 
These ring-like features suggest the possible stabilization of spiral spin liquid states, as proposed in other honeycomb materials, as well as the emergence of new magnetically ordered states upon including additional interactions. 
Further details regarding these phase diagrams will be explained in the following sections. Our results significantly update the phase diagrams of the model, and these results are important for future exploration of Kitaev candidate materials.

The rest of this paper is organized as follows.
In Sec.~\ref{subsec:model}, we introduce the Kitaev-Heisenberg-$\Gamma$ model. In Sec.~\ref{subsec:phase_diagram}, we briefly review the previous studies on its phase diagram. Our numerical methods are  described in Secs.~\ref{subsec:method_classical} and \ref{subsec:method_quantum}. We present results for the classical and quantum spins in Secs.~\ref{subsec:result_classical} and \ref{subsec:result_quantum}, respectively. In Sec.~\ref{subsec:additional}, we then discuss the effect of additional spin interactions on the states appearing, in highly frustrated regions, which are regarded as subdominant interactions in the materials.
Finally, we summarize our findings in Sec.~\ref{sec:summary}.

\section{Model and method}\label{sec:modelmethod}
\subsection{Model} \label{subsec:model}
In this study, we investigate phase diagrams of the Kitaev-Heisenberg-$\Gamma$ model~\cite{Rau2014, Rau2014a, Chaloupka2015} as a minimal model for candidate materials. The model is defined on a two-dimensional honeycomb lattice, and its Hamiltonian is given by
\begin{align}\label{eq:model}
    \mathcal{H}=\sum_{\gamma=x,y,z}\sum_{\langle i, j\rangle_\gamma}\biggl[J\mathbf{S}_i\cdot\mathbf{S}_j+K S_i^\gamma S_j^\gamma + \Gamma(S_i^\alpha S_j^\beta + S_i^\beta S_j^\alpha) \biggr],
\end{align}
where the summation of $\langle i,j\rangle_\gamma$ runs over pairs of nearest-neighbor sites $i$ and $j$ connected by $\gamma\ (=x,\ y,\ \text{or}\ z)$ bond, and $S^\gamma_i$ is the $\gamma$ component of the spin at site $i$: $\mathbf{S}_i=(S_i^x, S_i^y, S_i^z)$. The indices $\alpha$ and $\beta$ represent spin components that are distinct from each other and also from $\gamma$. In the classical case, $\mathbf{S}_i$ is treated as an $\mathrm{O}(3)$ vector, whereas in the quantum case it denotes an $S=1/2$ spin operator. The parameters $J$, $K$, and $\Gamma$ represent the coupling constants of the Heisenberg, Kitaev, and $\Gamma$ interactions, respectively, and they are parametrized as
\begin{align}\label{eq:coupling}
    J=\sin\theta\cos\phi, \quad K=\sin\theta\sin\phi, \quad \Gamma=\cos\theta,
\end{align}
by using two parameters $\theta\in[0,\ \pi]$ and $\phi\in[0,\ 2\pi)$, corresponding to the polar and azimuthal angles for the spherical coordinate. We take the energy unit as $\sqrt{J^2+K^2+\Gamma^2}=1$.
According to Eq.~\eqref{eq:coupling}, the northern hemisphere $\theta\in[0,\ \pi/2]$ and the southern hemisphere $\theta\in[\pi/2,\ \pi]$ correspond to the regions with $\Gamma\geq 0$ and $\Gamma\leq 0$, respectively. The Kitaev interaction is antiferromagnetic (AFM) ($K>0$) for $0 < \phi < \pi$, while it is ferromagnetic (FM) ($K<0$) for $\pi < \phi < 2\pi$. Meanwhile, the Heisenberg interaction is AFM ($J>0$) for $0\leq \phi <\pi/2$ and $3\pi/2 < \phi < 2\pi$, while FM ($J<0$) for $\pi/2 < \phi < 3\pi/2$. As pointed out in the previous study~\cite{Rau2014}, for classical spins, the parameters $(J, K, \Gamma)$ can be transformed into $(-J, -K, -\Gamma)$ by applying time reversal to the spins on one of the two sublattices of the honeycomb lattice. Hence, it is sufficient to consider only the region with $\Gamma \geq 0$. We set the nearest-neighbor bond length to unity.

In addition to the Heisenberg and $\Gamma$ interactions, the $\Gamma'$ interaction~\cite{Rau2014, Rau2014a, Chaloupka2015} --- another nearest-neighbor symmetric spin off-diagonal interaction --- and the third-neighbor Heisenberg interaction $J_3$~\cite{Kimchi2011, Sizyuk2014, Rusnacko2019} have also been proposed as subdominant interactions in candidate materials. The Hamiltonian of the $\Gamma'$ interaction is given by
\begin{align} \label{eq:Gamma_prime}
   \mathcal{H}_{\Gamma'}&=\Gamma' \sum_{\gamma=x,y,z}\sum_{\langle i, j\rangle_\gamma}\biggl[S_i^\alpha S_j^\gamma + S_i^\gamma S_j^\alpha + S_i^\beta S_j^\gamma + S_i^\gamma S_j^\beta \biggr],
\end{align}
where the summation and convention for $\alpha$, $\beta$, and $\gamma$ are the same as those in Eq.~\eqref{eq:model}, and $\Gamma'$ represents the coupling constant. Unlike the $\Gamma$ interaction represented by the last term in Eq.~\eqref{eq:model}, this interaction involves the $\gamma$ component of spins on the $\gamma$ bond, and it can become finite due to trigonal distortions of the edge-shared ligand octahedra in candidate materials~\cite{Rau2014a}. 
Furthermore, the Hamiltonian for the third-neighbor Heisenberg interaction is given by
\begin{align} \label{eq:J3}
   \mathcal{H}_{J_3}&=J_3 \sum_{\tripleangle{i, j}}\mathbf{S}_i\cdot\mathbf{S}_j,
\end{align}
where the summation $\tripleangle{i, j}$ runs over all pairs of third-neighbor sites $i$ and $j$, and $J_3$ denotes the coupling constant. This interaction has been suggested to play an important role in stabilizing the zigzag AFM order observed in many candidate materials~\cite{Kimchi2011}.
Although the primary focus of this paper is the Kitaev-Heisenberg-$\Gamma$ model, we discuss the effects of these additional interactions in Sec.~\ref{subsec:additional}.

\subsection{Previous studies on phase diagrams} \label{subsec:phase_diagram}
Shortly after the proposal by Jackeli and Khaliullin~\cite{Jackeli2009} the Kitaev-Heisenberg model without the $\Gamma$ interaction was proposed as a minimal model for Na$_2$IrO$_3$ and $\alpha$-Li$_2$IrO$_3$~\cite{Chaloupka2010, Chaloupka2013}. The ground-state phase diagram of this model has been extensively studied using Monte Carlo simulations for classical spins~\cite{Price2012, Price2013}, and, for $S=1/2$ quantum spins, by means of the exact diagonalization (ED)~\cite{Chaloupka2010, Chaloupka2013}, density matrix renormalization group (DMRG)~\cite{Jiang2011}, slave-particle mean-field~\cite{Schaffer2012}, tensor network~\cite{Iregui2014}, cluster mean-field~\cite{Gotfryd2017}, high-temperature expansion~\cite{Singh2017}, quantum Monte Carlo~\cite{Sato2021}, and PFFRG~\cite{Reuther2011a, Fukui2022a} methods.

Subsequently, it was found that the $\Gamma$ interaction plays an important role in materials such as $\alpha$-RuCl$_3$, which motivated studies of the ground-state phase diagram of the Kitaev-$\Gamma$ model without the Heisenberg interaction. The $\Gamma$ interaction renders the problem extremely challenging even at the classical level, due to the strong competition among a wide variety of magnetic orders, including coplanar and noncoplanar states with and without incommensurate modulations. 
Consequently, efforts to refine the phase diagram have continued in recent years using methods such as simulated annealing~\cite{Chern2020}, machine learning~\cite{Liu2019}, classical Monte Carlo~\cite{Rayyan2021, Chen2023}, and iterative variational minimization combined with classical Monte Carlo~\cite{Stavropoulos2024}.
The ground-state phase diagram in the quantum case has also been studied using various methods:  ED~\cite{Yamada2020, Rousochatzakis2020}, DMRG~\cite{Gohlke2018, Gordon2019, Gohlke2020}, tensor network~\cite{Lee2020}, and PFFRG~\cite{Buessen2021}. 

Regarding the Kitaev-Heisenberg-$\Gamma$ model considered in this paper, which incorporates both the Heisenberg and $\Gamma$ interactions into the Kitaev model, the magnetic phase diagram for classical spins was initially investigated in Ref.~\onlinecite{Rau2014} using the Luttinger-Tisza method~\cite{Luttinger1946, Litvin1974}. In parameter regions where the method loses validity and an exact ground state cannot be obtained [see Fig.~\ref{fig:main}(a)], the single-$Q$ ansatz minimization was employed; however, since these methods can only describe ordered states characterized by a single wave vector; this raises the possibility that the phase diagram may change significantly once multiple-$Q$ ordered states are taken into account. This expectation is also supported by previous studies of the Kitaev-$\Gamma$ model mentioned above, which proposed phase diagrams containing several multiple-$Q$ ordered phases. Subsequently the phase diagram was updated by using the simulated annealing method~\cite{Rau2014a} and the machine learning approach~\cite{Rao2021}, but much of the region still remains unclear. Crucially, since it is believed that many candidate materials fall within this region, clarification of the phase diagram of this region is extremely important.

For $S=1/2$ quantum spins, the ground-state phase diagram has been investigated using the ED method with a $24$-site cluster~\cite{Rau2014, Rau2014a, Rusnacko2019}, multiscale entanglement renormalization ansatz with a $24$-site cluster~\cite{Lou2015}, and the variational Monte Carlo method~\cite{Wang2019}. However, the large portions of the phase diagram also remain unclear, partly because large system sizes are required to properly address the multiple-$Q$ and incommensurate magnetic orders expected to emerge. In this paper we update the whole ground-state phase diagram for both classical and quantum spins by using state-of-the-art numerical methods described in the next two subsections.

\subsection{Numerical optimization method for classical spins} \label{subsec:method_classical}
In the study of the classical case, we treat $S=1/2$ spins as classical $\mathrm{O}(3)$ vectors with fixed length $1/2$. Each spin is parametrized as
\begin{align}
    \mathbf{S}_i(\vartheta_i, \varphi_i)=\frac{1}{2}\begin{pmatrix}
        \sin\vartheta_i\cos\varphi_i, & \sin\vartheta_i\sin\varphi_i, & \cos\vartheta_i
    \end{pmatrix},
\end{align}
where $\vartheta_i\in[0,\ \pi]$ and $\varphi_i\in[0,\ 2\pi)$ denote the polar and azimuthal angles, respectively. Note that these angles $(\vartheta_i, \varphi_i)$ are different from the angles ($\theta, \phi$) used to parametrize the coupling constants in Eq.~\eqref{eq:coupling}. 

To obtain the spin configurations for the lowest-energy state, we perform variational energy minimization by optimizing the directions of these classical $\mathrm{O}(3)$ vectors at each site using a gradient descent method. Specifically, we perform large-scale variational energy minimization based on an automatic differentiation scheme, using JAX~\cite{jax} and Optax~\cite{optax} libraries for machine learning~\cite{Pohle2023, Pohle2024, Ikegami2026a}. By utilizing GPUs, these libraries enable efficient optimization for large system sizes, allowing us to minimize bias and search for the lowest-energy states from a large number of random initial configurations. In the following calculations, we use Adam optimizer~\cite{adam}, and take at most $2000$ random seeds for each optimization flow.

We perform the optimization by progressively increasing the system size, starting from small clusters. At each step, noise is introduced to avoid trapping in local minima. 
Further details of the method are provided in Appendix~\ref{app:classical}.
A similar technique was used to study the classical phase diagram of the $S=3/2$ Kitaev-AKLT model~\cite{Ikegami2026a}.

The system sizes used in our main calculations are $N_{\mathrm{site}}=2\times L \times L$ with $L=36$ for the main results and $L=24$ for confirmation of the results, corresponding to $N_{\mathrm{site}}=2592$ and $N_{\mathrm{site}}=1152$, respectively. Here, $N_{\mathrm{site}}$ denotes the number of sites, i.e., the number of spins. 
To elucidate the phase diagram in detail, we perform the optimization on a grid of $281$ $\theta$ points and $1120$ $\phi$ points, corresponding to 314720 parameter points in total. 

To identify magnetic structures in each phase, we calculate the spin structure factor
\begin{align} \label{eq:Sq}
    \mathcal{S}(\mathbf{q})=\frac{1}{N_{\mathrm{site}}}\sum_{i, j}\mathrm{e}^{-\mathrm{i}\mathbf{q}\cdot(\mathbf{r}_i-\mathbf{r}_j)} \mathbf{S}_i\cdot\mathbf{S}_j,
\end{align}
where $\mathbf{r}_i$ and $\mathbf{r}_j$ represent the real-space coordinates of site $i$ and $j$, respectively. 
We also calculate the total scalar spin chirality defined as
\begin{align} \label{eq:chirality}
    \chi_{\mathrm{tot}}=\frac{1}{N_{\mathrm{site}}}\sum_{\mathbf{R}}\big[
        &\mathbf{S}_{\mathbf{R},\mathrm{A}}\cdot(\mathbf{S}_{\mathbf{R}+\mathbf{a}_1,\mathrm{A}} \times \mathbf{S}_{\mathbf{R}+\mathbf{a}_1+\mathbf{a}_2,\mathrm{A}})\notag\\
        +&\mathbf{S}_{\mathbf{R},\mathrm{A}}\cdot(\mathbf{S}_{\mathbf{R}+\mathbf{a}_1+\mathbf{a}_2,\mathrm{A}} \times \mathbf{S}_{\mathbf{R}+\mathbf{a}_2,\mathrm{A}})\notag\\
        +&\mathbf{S}_{\mathbf{R},\mathrm{B}}\cdot(\mathbf{S}_{\mathbf{R}+\mathbf{a}_1,\mathrm{B}} \times \mathbf{S}_{\mathbf{R}+\mathbf{a}_1+\mathbf{a}_2,\mathrm{B}})\notag\\
        +&\mathbf{S}_{\mathbf{R},\mathrm{B}}\cdot(\mathbf{S}_{\mathbf{R}+\mathbf{a}_1+\mathbf{a}_2,\mathrm{B}} \times \mathbf{S}_{\mathbf{R}+\mathbf{a}_2,\mathrm{B}})
        \big],
\end{align}
where $\mathrm{A}$ and $\mathrm{B}$ label the two sublattices of the honeycomb lattice. The summation runs over all unitcell of the lattice, $\mathbf{R}$. The vectors $\mathbf{a}_1=(\sqrt{3},\ 0)$ and $\mathbf{a}_2=(-\sqrt{3}/2,\ 3/2)$ represent the two primitive lattice vectors of the lattice. This quantity is used to identify phases with noncoplanar spin configurations.

\subsection{PFFRG method for quantum spins} \label{subsec:method_quantum}
To investigate the ground-state phase diagram for quantum spins, we adopt the PFFRG method~\cite{Reuther2010, Muller2024} in the zero-temperature limit, extended to models with spin off-diagonal interactions~\cite{Buessen2019}. This approach is a powerful framework for frustrated quantum spin systems, allowing calculations for large system sizes and enabling the identification of both quantum spin liquids and incommensurate magnetic orders, which are generally difficult to access by other methods. It has previously been applied to the Kitaev-Heisenberg model without the $\Gamma$ interaction~\cite{Reuther2011a, Fukui2022a} and to the Kitaev-$\Gamma$ model without the Heisenberg interaction~\cite{Buessen2021}.

In this method, the $S=1/2$ spin operator is expressed in terms of auxiliary fermions, also known as the Abrikosov fermions~\cite{Abrikosov1965}, as
\begin{align}\label{eq:pseudofermion}
    S^{\gamma}_i=\frac{1}{2}\sum_{\mu, \nu}f^\dagger_{i,\mu}\sigma^{\gamma}_{\mu\nu}f_{i,\nu},
\end{align}
where $f_{i,\mu}$ ($f^\dagger_{i,\mu}$) is an annihilation (creation) operator of the auxiliary fermion at site $i$ with spin projection $\mu\in\{\uparrow, \downarrow\}$, and $\sigma^{\gamma}$ is the $\gamma \ (=x, y, \mathrm{or}\ z)$ component of the Pauli matrices. We set the reduced Planck constant $\hbar$ to unity throughout this paper. Although this fermionic expression of spins enlarges the Hilbert space, the local constraint $\sum_{\mu}f^\dagger_{i,\mu}f_{i,\mu}=1$~\cite{Popov1988} is fulfilled automatically at zero temperature. 

By substituting Eq.~\eqref{eq:pseudofermion} into Eq.~\eqref{eq:model}, we obtain the fermionic Hamiltonian in the purely quartic form of the auxiliary fermion operators. To treat such a quartic Hamiltonian, we resort to the fermionic one-particle irreducible FRG theory~\cite{Salmhofer, Salmhofer2001, Kopietz, Metzner2012} which enables nonperturbative and unbiased calculations. Within this framework, we numerically solve the coupled integro-differential equation for the self-energy and the two-particle vertex function, i.e., the renormalized dynamical couplings between auxiliary fermions. To incorporate quantum fluctuations sufficiently to describe quantum spin liquid states, we adopt the Katanin truncation scheme~\cite{Katanin2004}, where the infinite hierarchy of flow equations is truncated at the one-loop level in a fully self-consistent manner. 

From the obtained self-energy and two-particle vertex function, we calculate the $\gamma$ component of the Fourier transformed static spin susceptibility $\chi^{\gamma\gamma,\Lambda}(\mathbf{q})$ at momentum $\mathbf{q}$ defined as
\begin{align}
    \chi^{\gamma\gamma,\Lambda}(\mathbf{q})=\frac{1}{N_{\mathrm{site}}}\sum_{i, j}\mathrm{e}^{-\mathrm{i}\mathbf{q}\cdot(\mathbf{r}_i-\mathbf{r}_j)}\chi^{\gamma\gamma, \Lambda}_{ij},
\end{align}
where $\chi^{\gamma\gamma, \Lambda}_{ij}$ is the static spin susceptibility in real space calculated as
\begin{align}
    \chi^{\gamma\gamma,\Lambda}_{ij}=\int^{\infty}_0\mathrm{d}\tau\ \langle T_{\tau} S^{\gamma}_i(\tau)S^{\gamma}_j(0)\rangle_{\Lambda}.
\end{align}
Here, $S^{\gamma}_i(\tau)=\mathrm{e}^{\tau\mathcal{H}}S^{\gamma}_i\mathrm{e}^{-\tau\mathcal{H}}$ with the imaginary time $\tau$, and $\langle T_\tau \cdots\rangle_{\Lambda}$ means the expectation value of imaginary-time-ordered operators at the cutoff scale $\Lambda$. Owing to $C_3$ symmetry of the model, the total spin susceptibility defined as $\chi^{\Lambda}(\mathbf{q})=\chi^{xx,\Lambda}(\mathbf{q})+\chi^{yy,\Lambda}(\mathbf{q})+\chi^{zz,\Lambda}(\mathbf{q})$ has six-fold symmetry in momentum space. The parameter $\Lambda$ represents the cutoff energy scale of the renormalization group calculation: the low-energy effective theory is obtained by successively integrating out states with energy higher than $\Lambda$. A magnetic instability is signaled by divergence of $\chi^{\Lambda}(\mathbf{q})$ at a momentum corresponding to the ordering vector, and the corresponding cutoff scale $\Lambda$ is referred to as the critical cutoff energy scale $\Lambda_{\mathrm{c}}$. In practice, however, due to the finite system size and the finite frequency grid, the $\Lambda$ dependence of $\chi^{\Lambda}(\mathbf{q})$ shows a kink or cusp rather than a true divergence. Thus, we use such anomaly to identify magnetic instabilities. When $\Lambda_{\mathrm{c}}$ is sufficiently larger than the characteristic magnetic energy scale, it can be regarded as an estimate of the transition temperature $T_{\mathrm{c}}$ according to $T_{\mathrm{c}} \simeq (\pi/2)\Lambda_{\mathrm{c}}$~\cite{Iqbal2016, Buessen2016, Buessen2019}. In contrast, the absence of such anomaly down to $\Lambda \to 0$ suggests that the system realizes a QSL ground state, without showing any magnetic instability. 

In the following calculations, we take the zero-temperature limit as is customary in PFFRG calculations, and use the logarithmic frequency mesh with 64 positive Matsubara frequency points between $10^{-4}$ and $50$. We also adopt the logarithmic $\Lambda$ grid starting from $\Lambda_{\mathrm{max}}=500$ to $\Lambda_{\mathrm{min}}\simeq 10^{-1}$ by multiplying a factor of $0.98$ sequentially. We include two-particle vertex functions between two sites up to $12$th neighbors, which corresponds to a finite-size cluster with $N_{\mathrm{site}}=235$. To elucidate the phase diagram, we carry out calculations at more than 720 parameter points as indicated by the symbols in Figs.~\ref{fig:main}(b) and \ref{fig:main}(c), in addition to the data from previous studies of the Kitaev-Heisenberg model~\cite{Fukui2022a}. Further details of the PFFRG method used here can be found in Ref.~\onlinecite{Fukui2022a}.

\section{Results and Discussion} \label{sec:result}
\subsection{Classical case} \label{subsec:result_classical}

\begin{figure}
    \centering
     \includegraphics[width=\linewidth,clip]{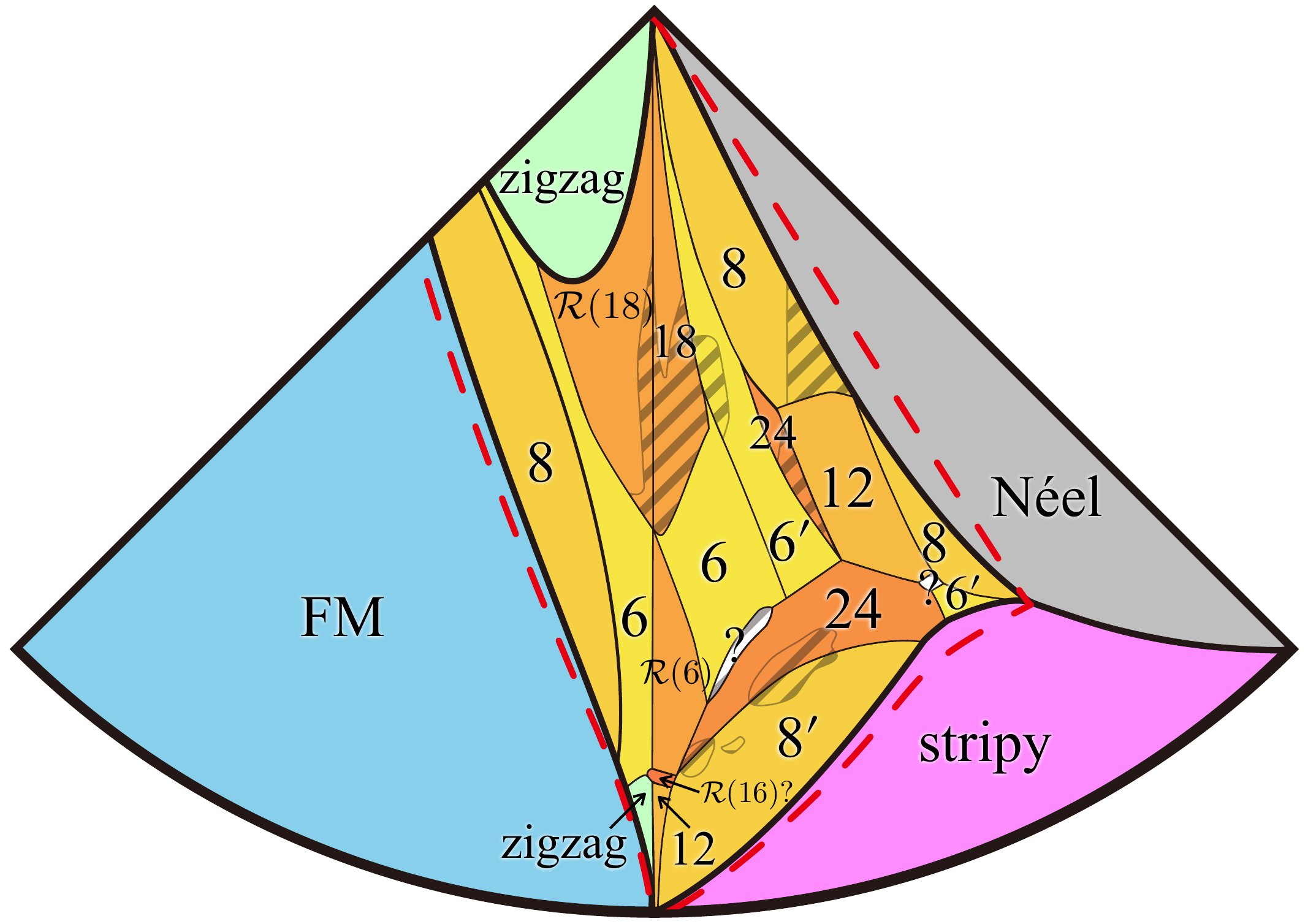}
    \caption{Enlarged view of the lower quadrant of Fig.~\ref{fig:main}(a). The notations are common to those in Fig.~\ref{fig:main}(a).
    }
    \label{fig:classical_phase}
\end{figure}

\begin{figure}
    \centering
     \includegraphics[width=0.8\linewidth,clip]{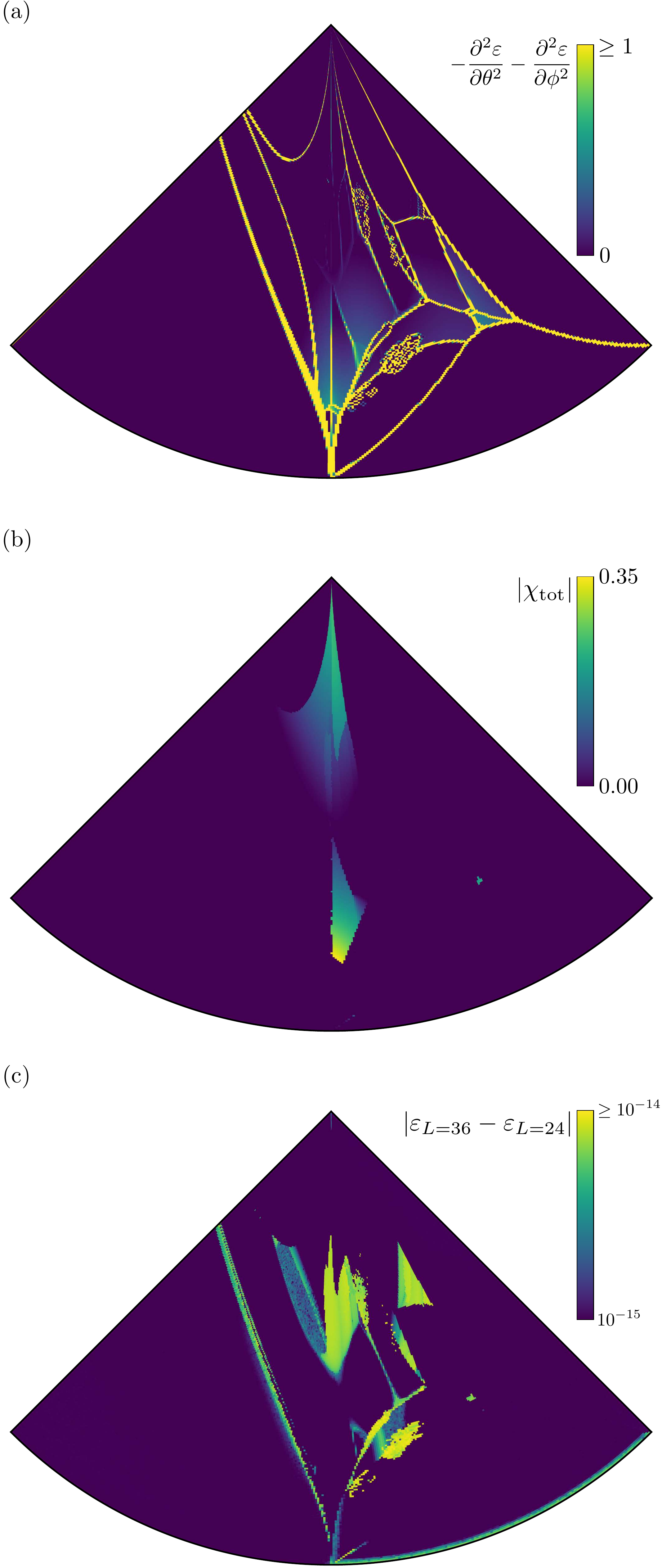}
    \caption{(a) Second-order derivative, (b) absolute value of the total scalar spin chirality $\chi_{\mathrm{tot}}$, and (c) system size dependence of the energy per site $\varepsilon$, calculated from the optimized spin configurations. As in Fig.~\ref{fig:classical_phase}, only the lower quadrant of the phase diagram is shown.
    }
    \label{fig:classical_data}
\end{figure}

\begin{figure}
    \centering
     \includegraphics[width=0.9\linewidth,clip]{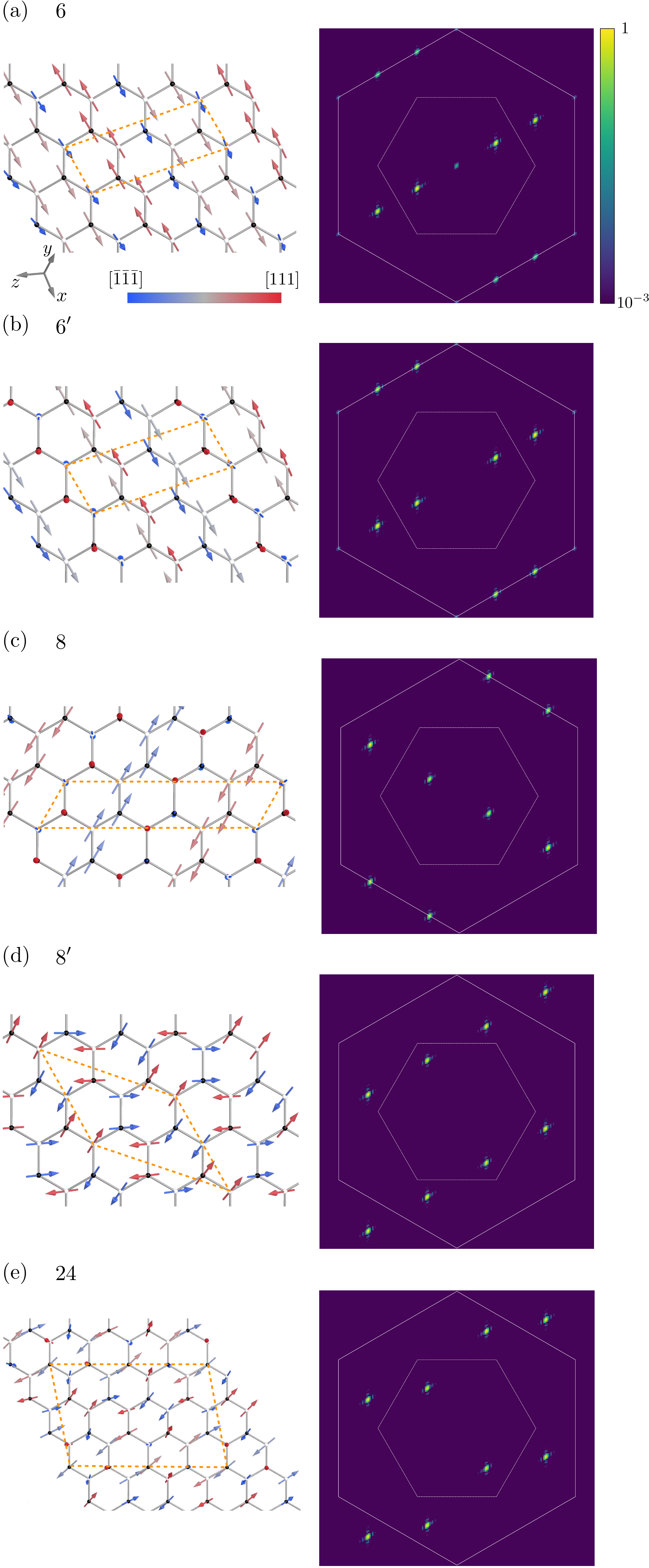}
    \caption{Typical data of phases (a) $6$, (b) $6'$, (c) $8$, (d) $8'$, and (e) $24$. The left panels show the optimized spin configurations. The colors of the arrows encode the magnitude of the out-of-plane component: $[111]$ and $[\bar{1}\bar{1}\bar{1}]$ directions are shown in red and blue, respectively. Gray represents the in-plane component. The color code and spin axes are shown in (a). Orange dotted lines indicate the magnetic unit cells. The right panels show $\mathbf{q}$ dependence of $\mathcal{S}(\mathbf{q})/\mathrm{max}[\mathcal{S}(\mathbf{q})]$ in logarithmic scale. The color code is shown in (a). The inner white dotted and outer white solid lines represent the first Brillouin zone and the enlarged zone including up to the third one, respectively. 
    }
    \label{fig:jax_spin}
\end{figure}

Figure~\ref{fig:main}(a) shows the phase diagram in the classical case obtained from our numerical minimization. As noted in Sec.~\ref{subsec:model}, for classical spins, it is sufficient to determine the phase diagram only for $\Gamma \geq 0$. We obtain dominant phases, FM, N\'eel AFM, zigzag, stripy, and vortex (also called 120$^\circ$), in agreement with previous studies~\cite{Rau2014, Rau2014a}.
However, in the regions where the Luttinger-Tisza method is no longer valid [enclosed by the red dashed line in Fig.~\ref{fig:main}(a)], we identify highly competitive and complicated phase structures that are remarkably different from those reported in the previous studies.
To discuss this region in detail, Fig.~\ref{fig:classical_phase} shows an enlarged view of the lower quadrant ($5/4 \leq \phi/\pi \leq 7/4$) of the phase diagram. We find that, in the entire parameter region where the Luttinger-Tisza method fails, numerous coplanar and noncoplanar orders with or without incommensurate modulations appear, resulting in a \textit{zoo of noncollinear orders}. Such a diverse set of competing magnetic phases can be attributed to strong frustration driven by the competition among three magnetic interactions $J$, $K$, and $\Gamma$. 

In this region, we identify nine types of magnetically ordered phases: two coplanar six-site ordered phases, denoted by $6$ and $6'$; two coplanar eight-site ordered phases, denoted by $8$ and $8'$; one  coplanar 12-site ordered phase, denoted by $12$; three noncoplanar 18-site ordered phases, denoted by $18$, $\mathcal{R}(18)$, and $\mathcal{R}(6)$; and one coplanar 24-site ordered phase, denoted by $24$. We also observe an ordered phase, denoted by $\mathcal{R}(16)$, that closely resembles the noncoplanar 48-site ordered phase proposed in a previous study~\cite{Stavropoulos2024}, along with two additional ordered phases denoted by ``?''.
However, none of these phases exhibit sufficient convergence to allow definitive identification. The hatched regions indicate the presence of incommensurate modulations. 
The vertical line in the center corresponds to the Kitaev-$\Gamma$ model with $0\leq \theta/\pi \leq 1/2$ and $\phi/\pi=3/2$, and the phase boundaries along this line in our result agree well with the latest updated phase diagram of the Kitaev-$\Gamma$ model in Ref.~\onlinecite{Stavropoulos2024}. 
For the ordered phases previously reported in the Kitaev-$\Gamma$ model, we follow the nomenclature of Ref.~\onlinecite{Stavropoulos2024}, except for the phase labeled $6$, which corresponds to $6'$ in that reference. 
Note that all of these additional magnetic orders are noncollinear. We will discuss these ordered phases in more detail later. 

The data used to identify these phases and the phase boundaries between them are shown in Fig.~\ref{fig:classical_data}. The phase boundaries are primarily determined from the second-order derivative of the optimized energy per site $\varepsilon$ with respect to $\theta$ and $\phi$ shown in Fig.~\ref{fig:classical_data}(a). The nature of each phase is identified by analyzing the optimized spin configurations and calculating the spin structure factor in Eq.~\eqref{eq:Sq}. In addition, noncoplanarity of the spin configuration is detected through a nonzero scalar spin chirality defined in Eq.~\eqref{eq:chirality}, as shown in Fig.~\ref{fig:classical_data}(b).
To further classify the states as collinear, coplanar, and noncoplanar, we analyze the eigenvalues of the spin inertia tensor~\cite{Sklan2013}, as presented in Appendix~\ref{app:inertia}.

In Fig.~\ref{fig:classical_data}(a), several regions show relatively large energy derivatives accompanied by noisy behavior. These regions are likely to host incommensurate order. To identify such incommensurate regions, we evaluate the energy difference between clusters with $L=24$ and $36$, as shown in Fig.~\ref{fig:classical_data}(c). A strong system-size dependence of the energy is taken as a criterion for identifying incommensurate phases. In regions where the  system-size dependence is finite but weak, we determine the presence or absence of incommensurate order by directly examining the optimized spin configurations and the corresponding structure factors at several representative parameter points. Based on these analyses, we identify the hatched regions in Figs.~\ref{fig:main}(a) and \ref{fig:classical_phase}. 

Let us discuss in detail each of the noncollinear magnetic orders that appear in the phase diagram shown in Fig.~\ref{fig:classical_phase}. We first briefly review the magnetic ordered phases that have already been identified in previous studies of the Kitaev-$\Gamma$ model~\cite{Stavropoulos2024}. Owing to the presence of the six-sublattice transformation $\mathcal{R}$~\cite{Rousochatzakis2015a, Rousochatzakis2024, Stavropoulos2024}, the ground states of the Kitaev-$\Gamma$ model are always doubly degenerate. Starting from the ferromagnetic Kitaev limit and increasing $\Gamma>0$, the system successively realizes degenerated pairs of ordered states, namely zigzag and $12=\mathcal{R}(\mathrm{zigzag})$, $16$ and $\mathcal{R}(16)$, $6$ and $\mathcal{R}(6)$, and $18$ and $\mathcal{R}(18)$. In our optimization calculations, one of the two degenerate states is obtained depending on initial random configurations.

In the presence of a finite Heisenberg interaction $J$, however, the six-sublattice transformation is no longer valid. As a consequence, for positive $J$, the system chooses the phases $12$, $\mathcal{R}(6)$, and $18$, whereas for negative $J$, the phases zigzag, $6$, and $\mathcal{R}(18)$ appear.
As mentioned above, our results reveal a tiny region close to the $48$-site ordered phase $\mathcal{R}(16)$, although full convergence is not achieved. In contrast, we do not identify any regions corresponding to the $16$-site ordered phase, denoted phase $16$ in the previous study on the Kitaev-$\Gamma$ model~\cite{Stavropoulos2024}. This may indicate strong fragility of this $16$-site ordered phase of the Kitaev-$\Gamma$ model against the Heisenberg interaction.

We further identify four additional ordered phases --- $6'$, $8$, $8'$, and $24$ --- that emerge due to the effect of the Heisenberg interaction. As noted above, all of these correspond to coplanar magnetic orders. Figure~\ref{fig:jax_spin} presents schematic figures of the spin configurations and the corresponding spin structure factors for these newly identified phases. For comparison, we also include data for the phase $6$ in order to highlight the difference from the phase $6'$. The spin structure factors of the phases $6$, $6'$, and $8$ exhibit strong similarities up to $C_3$ symmetry of the model. Among them, only the phase $6$ shows a peak at $\mathbf{q}=\mathbf{0}$, indicating the presence of a finite net magnetization. Although their real-space periods are distinct, the spin pinning directions are also largely similar up to a $C_3$ rotation. These phases are adjacent to one another in the phase diagram in Figs.~\ref{fig:main}(a) and \ref{fig:classical_phase}. By contrast, the phases $8'$ and $24$ display similar spin structure factors, and their spin pinning directions also appear to be closely related. These two phases are likewise adjacent in the phase diagram.

In summary, our classical analysis substantially refines the phase diagram in the regions where the Luttinger-Tisza method breaks down, and provides a significantly improved description of these regions. We find that they exhibit a highly intricate phase structure characterized by strong competition among numerous noncollinear orders, forming what may be described as a \textit{zoo of noncollinear orders}. 
Through careful analyses of the optimized spin configurations, spin structure factors, scalar spin chirality, and system-size dependence of the energy, we identify and characterize the individual phases and identify four new coplanar ordered phases, $6'$, $8$, $8'$, and $24$.

\subsection{Quantum case} \label{subsec:result_quantum}

\begin{figure}
    \centering
     \includegraphics[width=0.9\linewidth,clip]{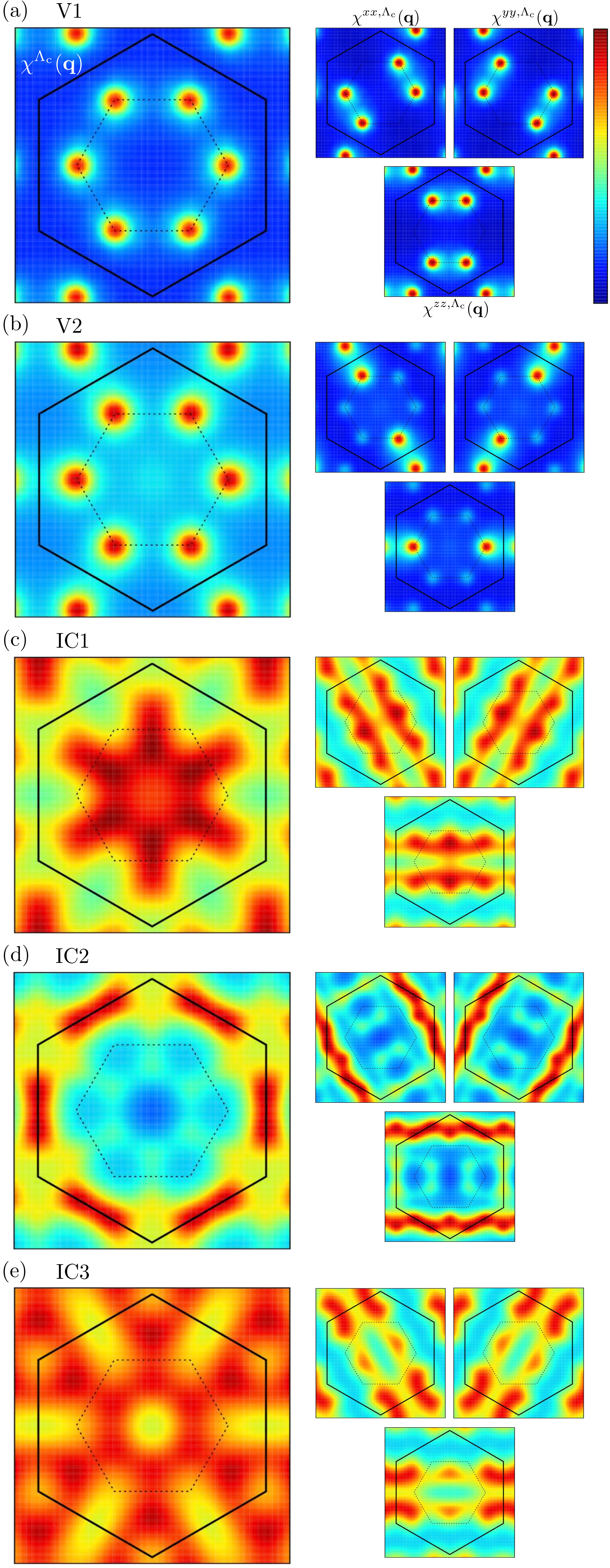}
    \caption{
        $\mathbf{q}$ dependence of the spin susceptibilities for (a) $\mathrm{V1}$, (b) $\mathrm{V2}$, (c) $\mathrm{IC1}$, (d) $\mathrm{IC2}$, and (e) $\mathrm{IC3}$ states. The left panels show $\chi^{\Lambda_{\mathrm{c}}}(\mathbf{q})$, and the right small panels show $\chi^{xx, \Lambda_{\mathrm{c}}}(\mathbf{q})$, $\chi^{yy, \Lambda_{\mathrm{c}}}(\mathbf{q})$, and $\chi^{zz, \Lambda_{\mathrm{c}}}(\mathbf{q})$. The color code is normalized within each plot. The inner black dotted and outer black solid lines represent the first Brillouin zone and the enlarged zone including up to the third one, respectively.
    }
    \label{fig:pffrg_chiq}
\end{figure}

\begin{figure}
    \centering
     \includegraphics[width=\linewidth,clip]{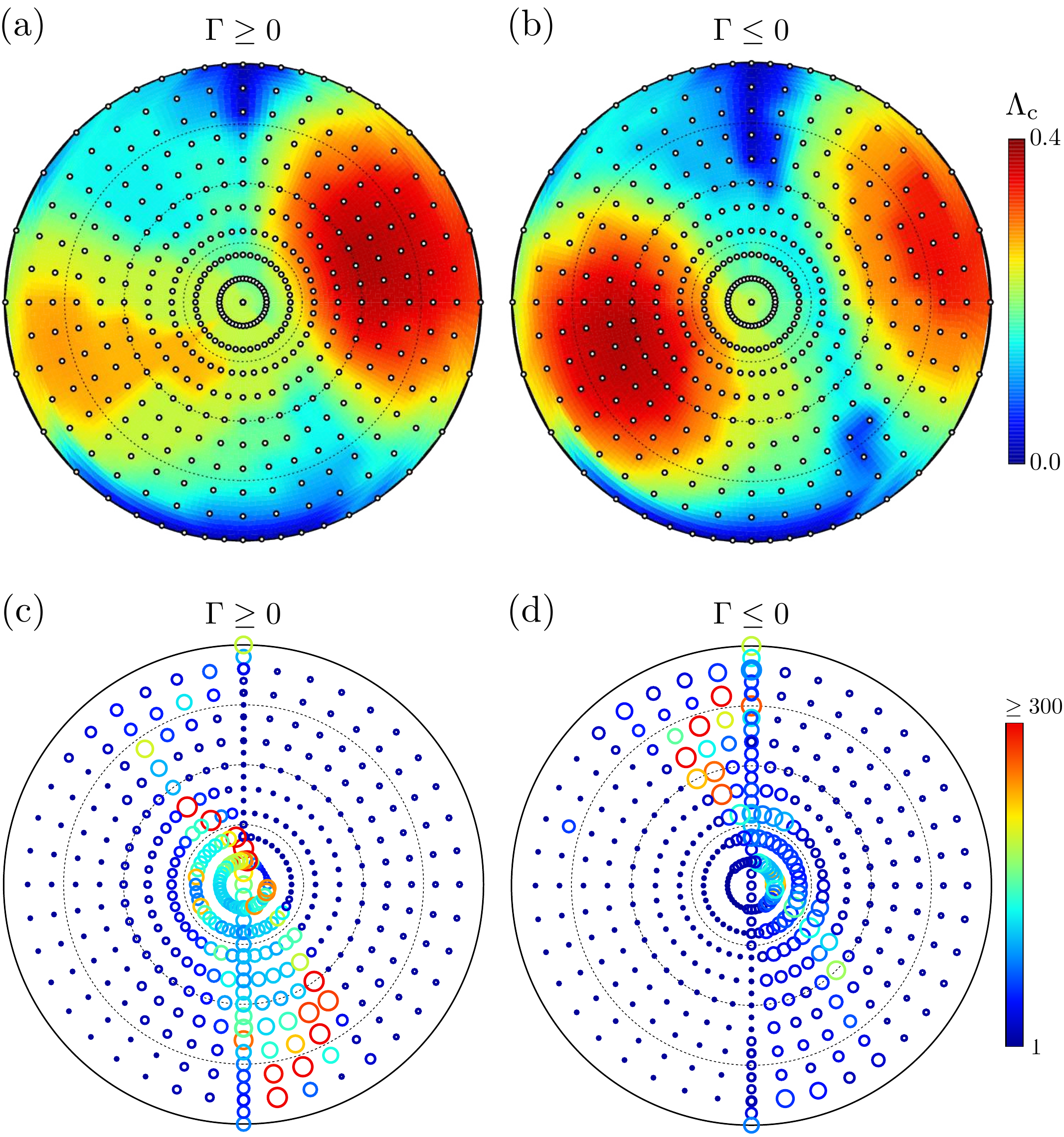}
    \caption{Critical cutoff energy scale $\Lambda_{\mathrm{c}}$ for (a) $\Gamma \geq 0$ and (b) $\Gamma \leq 0$.
    The white circles denote the parameter points at which the calculations were performed, while the colormap is obtained by interpolating between these data points. The data along the outer circle are taken from a previous study of the Kitaev-Heisenberg model~\cite{Fukui2022a}, with rescaling applied to match the energy scale of the present calculations.
    (c) and (d) show the approximate measure of degeneracy (see main text) for $\Gamma \geq 0$ and $\Gamma \leq 0$, respectively. The radius and outline color of each circle are proportional to the corresponding degeneracy.
    }
    \label{fig:degen_data}
\end{figure}

\begin{figure}
    \centering
     \includegraphics[width=\linewidth,clip]{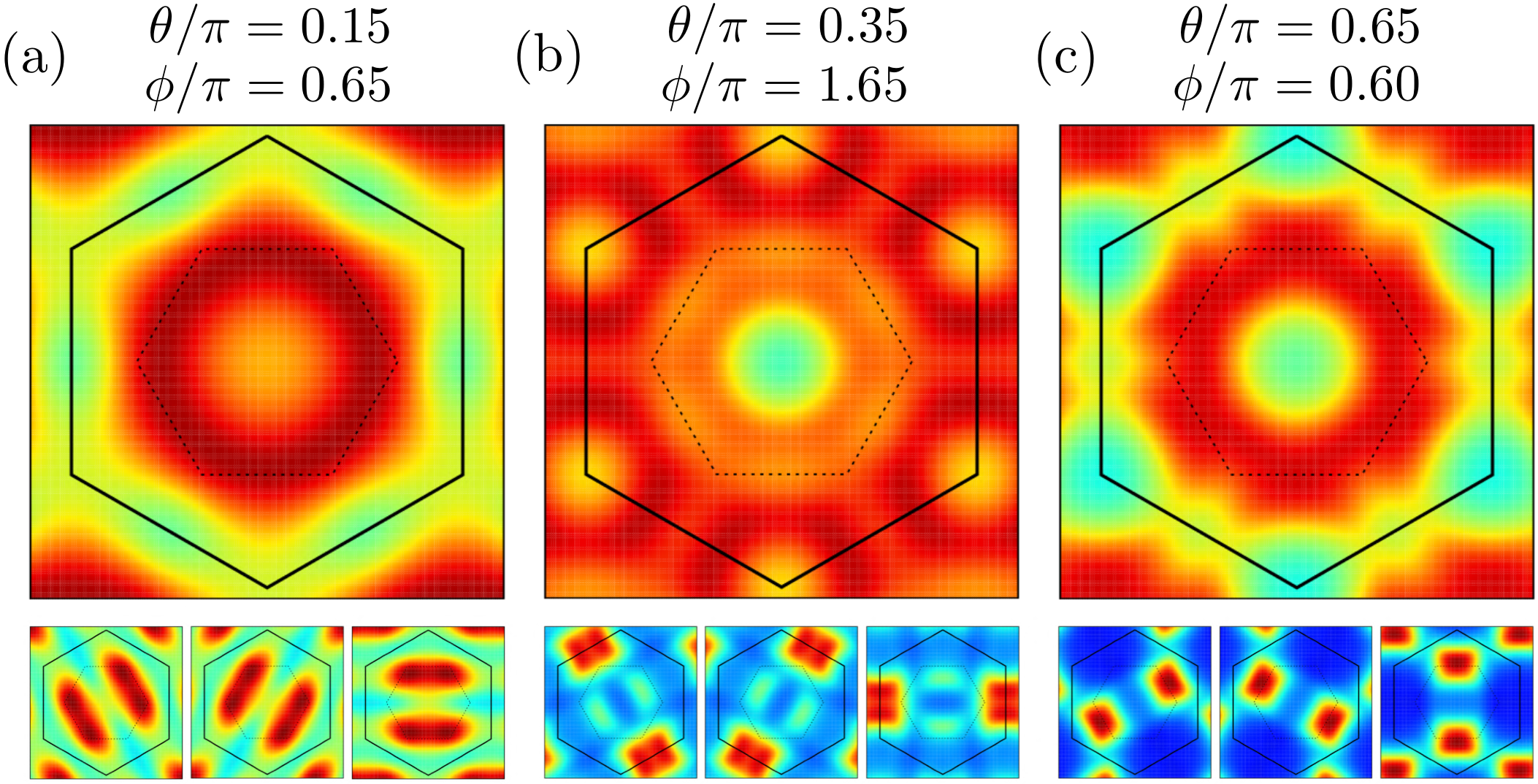}
    \caption{Typical $\mathbf{q}$ dependences of $\chi^{\Lambda_{\mathrm{c}}}(\mathbf{q})$ in the regions denoted ``highly frustrated'' in Figs.~\ref{fig:main}(b) and \ref{fig:main}(c). The data shown in (a), (b), and (c) are obtained at $(\theta/\pi,\ \phi/\pi)=(0.15,\ 0.65)$, $(0.35,\ 1.65)$, and $(0.65,\ 0.60)$, respectively. Bottom panels show $\chi^{xx,\Lambda_{\mathrm{c}}}(\mathbf{q})$, $\chi^{yy, \Lambda_{\mathrm{c}}}(\mathbf{q})$, and $\chi^{zz,\Lambda_{\mathrm{c}}}(\mathbf{q})$, from left to right.
    The notations are common to those in Fig.~\ref{fig:pffrg_chiq}.
    }
    \label{fig:degen_chiq}
\end{figure}

We next discuss the results for the quantum spin case. As briefly mentioned in Sec.~\ref{sec:intro}, Figs.~\ref{fig:main}(b) and \ref{fig:main}(c) show the ground-state phase diagrams obtained by the PFFRG method for $\Gamma \geq 0$ and $\Gamma \leq 0$, respectively. In contrast to the classical case, the quantum model does not possess a transformation relating $(J, K, \Gamma)$ to $(-J, -K, -\Gamma)$, and therefore both positive and negative values of $\Gamma$ must be considered separately. Indeed, a direct comparison of Figs.~\ref{fig:main}(b) and \ref{fig:main}(c) reveals that the phase boundaries do not correspond to one another.

First, similarly to the classical phase diagram in Fig.~\ref{fig:main}(a) and the previous ED studies~\cite{Rau2014, Rau2014a, Rusnacko2019}, the regions in which the Luttinger-Tisza method remains valid in the classical limit [outside the region enclosed by the red dashed line in Fig.~\ref{fig:main}(a)] exhibit conventional magnetic orders such as FM, N\'eel, zigzag, stripy, and vortex ($120^\circ$) phases. Note that, for $\Gamma < 0$, the region where the Luttinger-Tisza method breaks down is inverted with respect to the center of the circle, relative to that shown in Fig.~1(a) for $\Gamma > 0$. Since different vortex orders are realized for $\Gamma>0$ and $\Gamma<0$, we distinguish them here as $\mathrm{V1}$ and $\mathrm{V2}$, respectively, following a previous PFFRG study on the Kitaev-$\Gamma$ model~\cite{Buessen2021}.

Beyond these conventional ordered phases, the quantum system exhibits features absent in the classical case. In particular, we identify extended QSL phases in the vicinity of the FM ($K=-1,\ J=\Gamma=0$) and AFM ($K=1,\ J=\Gamma=0$) Kitaev models, which correspond to $(\theta/\pi,\ \phi/\pi)=(1/2,\ 3/2)$ and $(1/2,\ 1/2)$, respectively.
Consistent with the previous studies~\cite{Rau2014, Rau2014a}, we find that the FM Kitaev QSL in the vicinity of the FM Kitaev model is more robust against the Heisenberg interaction than its AFM counterpart, while it is more fragile against the $\Gamma$ interaction. Conversely, the AFM Kitaev QSL is more robust against the $\Gamma$ interaction and is stabilized over a wider parameter range in the $\theta$ direction. Moreover, it is slightly more stable for negative $\Gamma$ than for positive $\Gamma$.

Furthermore, quantum fluctuations also have a strong impact on magnetically ordered phases, particularly in regions where the Luttinger-Tisza method breaks down in the classical limit. The \textit{zoo of noncollinear orders} identified in the classical case is largely replaced by a small number of incommensurate ordered phases in the quantum system, denoted $\mathrm{IC1}$, $\mathrm{IC2}$, and $\mathrm{IC3}$. As with $\mathrm{V1}$ and $\mathrm{V2}$, the nomenclature for $\mathrm{IC1}$ and $\mathrm{IC2}$ follows Ref.~\cite{Buessen2021}.
By contrast, $\mathrm{IC3}$ denotes a distinct incommensurate magnetic order that does not appear in the Kitaev-$\Gamma$ model and is newly identified in the present study. 
The corresponding static spin susceptibility obtained by the PFFRG method for these magnetic orders is shown in Fig.~\ref{fig:pffrg_chiq}. The data for the phases common to the Kitaev-Heisenberg model including the Kitaev QSL phase have been already presented in Ref.~\onlinecite{Fukui2022a}. 
The susceptibility profile of the $\mathrm{IC3}$ phase is intermediate between those of the $\mathrm{IC1}$ phase and the stripy or N\'eel phases, suggesting that the $\mathrm{IC3}$ phase is stabilized by the competition among these ordering tendencies. Indeed, the $\mathrm{IC3}$ phase emerges only when the Heisenberg interaction drives the system away from the $\mathrm{IC1}$ state toward the stripy or N\'eel states in the phase diagram shown in Fig.~\ref{fig:main}(b). This highlights the importance of the antiferromagnetic Heisenberg interaction and, at the same time, explains why the $\mathrm{IC3}$ phase was not observed in the previous study of the Kitaev-$\Gamma$ model~\cite{Buessen2021}. We also note that the $\mathrm{IC1}$ phase appears near the parameter region where the zigzag order is stabilized in the classical case. Consistently, the susceptibility profile of the $\mathrm{IC1}$ phase remains very similar to that of the zigzag phase.

In addition, we identify several subsidiary ordered phases beyond the primary magnetic phases discussed above: FM$^*$, Neel$^*$, $\mathrm{IC}'$, $\mathrm{V2}'$, and $\mathrm{stripy}'$. 
The FM$^*$ and N\'eel$^*$ phases exhibit the same peak positions in the total spin susceptibility $\chi^{\Lambda}(\mathbf{q})$ as the FM and N\'eel phases, respectively. However, their component-resolved susceptibilities $\chi^{\gamma\gamma,\Lambda}(\mathbf{q})$ develop peaks at wave vectors shifted from those of the corresponding conventional orders. Both phases appear near phase boundaries in regions where the $\Gamma$ interaction is relatively strong.
The $\mathrm{IC'}$ phase emerges as an intermediate phase between the $\mathrm{IC}1$ and $\mathrm{IC}2$ phases. The peak positions of $\chi^{\Lambda}(\mathbf{q})$ differ from those of both the $\mathrm{IC}1$ and $\mathrm{IC}2$ phases. 
The $\mathrm{V2}'$ phase corresponds to an ordered state in which the susceptibility peaks are slightly shifted from those of the $\mathrm{V2}$ phase toward incommensurate wave vectors.
The stripy$'$ phase exhibits peak positions in $\chi^{\Lambda}(\mathbf{q})$ similar to those of the stripy phase, while its $\chi^{\gamma\gamma, \Lambda}(\mathbf{q})$ show dominant peaks along different edges of the Brillouin zone compared to the stripy order. This phase appears in regions where $\Gamma$ is large and negative.
The corresponding susceptibility data for these phases are provided in Appendix~\ref{app:pffrg}.

We also identify regions that remain strongly frustrated, labeled ``highly frustrated'' in Figs.~\ref{fig:main}(b) and \ref{fig:main}(c) and enclosed by black dashed lines. In these regions, $\Lambda_{\rm c}$ is strongly suppressed, as shown in Figs. 6(a) and 6(b), indicating that competing interactions inhibit magnetic ordering tendencies.
To further quantify the degree of frustration in these regions, we characterize the multi-peak structure, or flatness, of the $\mathbf{q}$-dependence of $\chi^{\Lambda_{\rm c}}(\mathbf{q})$. Specifically, we introduce an approximate measure of degeneracy defined as the number of wave vectors satisfying $\chi^{\Lambda_{\mathrm{c}}}(\mathbf{q}) \geq 0.99\times\chi^{\Lambda_{\mathrm{c}}}(\mathbf{q}_{\mathrm{max}})$, where $\mathbf{q}_{\mathrm{max}}$ is the wave vector at which $\chi^{\Lambda_{\mathrm{c}}}(\mathbf{q})$ attains its maximum. In other words, we count the number of momentum points whose susceptibility exceeds $99\%$ of its peak value. The resulting values are shown in Fig.~\ref{fig:degen_data}.
Combining these data with the values of $\Lambda_{\rm c}$ in Figs.~\ref{fig:degen_data}(a) and \ref{fig:degen_data}(b), we identify three regions where $\Lambda_{\mathrm{c}}$ is suppressed and the degeneracy is large, which are denoted ``highly frustrated'' in Figs.~\ref{fig:main}(b) and \ref{fig:main}(c). 

We emphasize that although $\Lambda_{\rm c}$ is strongly suppressed in these highly frustrated regions, it remains finite, implying that these states are not fully disordered but still retain magnetic ordering tendencies.
Typical data for $\chi^{\Lambda}(\mathbf{q})$ in these regions are presented in Fig.~\ref{fig:degen_chiq}.
The total susceptibility exhibits a ring-like structure in momentum space, closely resembling that observed in spiral spin liquid states~\cite{Bergman2007, Yao2021} reported in other honeycomb-lattice systems, such as the van der Waals honeycomb magnet FeCl$_3$~\cite{Gao2022} and the $J_1$-$J_2$ Heisenberg model on the honeycomb lattice~\cite{Okumura2010, Baez2017}.
Unlike conventional spiral spin liquid behavior, this ring-like structure emerges only after summing over the spin components. This feature can be naturally attributed to the bond-dependent anisotropic interactions, which favor different ordering vectors for different spin components. 
We note, however, that the susceptibility maxima in the individual spin components remain rather broad, indicating that no single ordering wave vector is strongly selected. This reflects the highly frustrated nature of this parameter region.
Within the results in the present study, $\Lambda_{\mathrm{c}}$ is not sufficiently suppressed to realize spiral spin liquid states in these highly frustrated regions: the fine structures in Figs.~\ref{fig:degen_chiq}(a) and \ref{fig:degen_chiq}(c) suggest a tendency toward $\mathrm{IC1}$, while that in Fig.~\ref{fig:degen_chiq}(b) indicates $\mathrm{IC3}$. Nevertheless, additional interactions, such as ring-exchange interactions, may further enhance quantum fluctuations and potentially stabilize spiral spin liquid states.

In summary, quantum fluctuations strongly reshape the phase diagram compared with the classical case. The numerous competing noncollinear orders found classically are largely suppressed and replaced by a limited number of incommensurate phases, including the newly identified $\mathrm{IC3}$ phase, together with extended highly frustrated regions. In these regions, the critical scale is strongly suppressed and the spin susceptibility exhibits ring-like structures in momentum space, indicating proximity to spiral spin liquid behavior. Although fully developed spiral spin liquid states are not stabilized in the present model, these regions may host novel magnetic orders or spiral spin liquid phases when additional interactions are included.

\subsection{Effect of additional interactions} \label{subsec:additional}

\begin{figure}
    \centering
     \includegraphics[width=0.9\linewidth,clip]{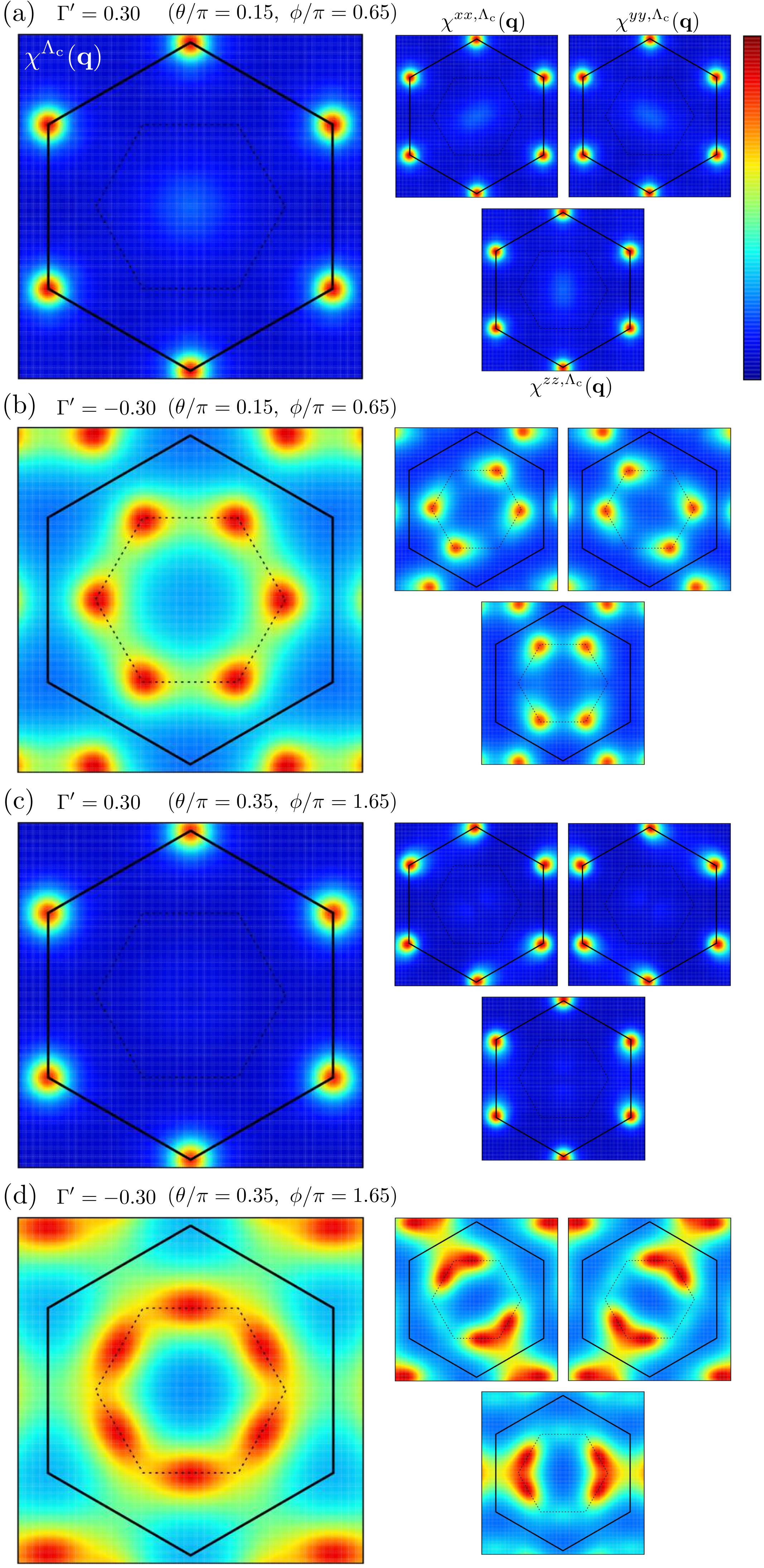}
    \caption{$\mathbf{q}$ dependences of the spin susceptibilities for the Kitaev-Heisenberg-$\Gamma$ model with the $\Gamma'$ interaction at (a) and (b) $(\theta/\pi,\ \phi/\pi)=(0.15,\ 0.65)$, and (c) and (d) $(0.35,\ 1.65)$. The parameter $\Gamma'$ is set to $0.30$ in (a) and (c), and to $-0.30$ in (b) and (d). 
    The notations are common to those in Fig.~\ref{fig:pffrg_chiq}.
    }
    \label{fig:Gprime}
\end{figure}

\begin{figure}
    \centering
     \includegraphics[width=0.9\linewidth,clip]{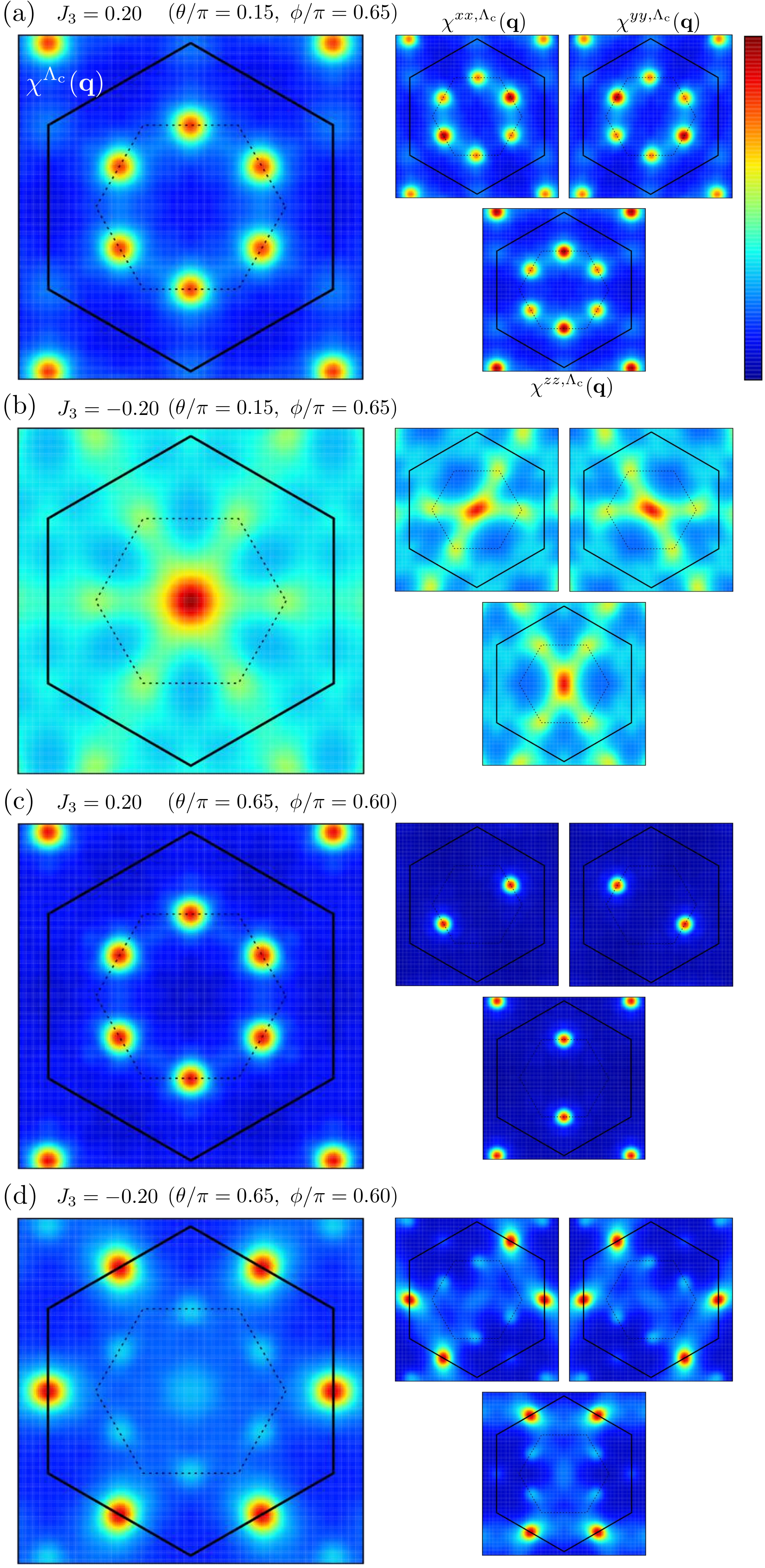}
    \caption{$\mathbf{q}$ dependences of the spin susceptibilities for the Kitaev-Heisenberg-$\Gamma$ model with the third-neighbor Heisenberg interaction, $J_3$, at (a) and (b) $(\theta/\pi,\ \phi/\pi)=(0.15,\ 0.65)$, and (c) and (d) $(0.65,\ 0.60)$. The parameter $J_3$ is set to $0.20$ in (a) and (c), and to $-0.20$ in (b) and (d). 
    The notations are common to those in Fig.~\ref{fig:pffrg_chiq}. 
    }
    \label{fig:J3}
\end{figure}

The susceptibility features indicative of strong degeneracy also implies that, in addition to the possibility of spiral spin liquids, these systems are susceptible to transitions into new magnetic orders which do not appear in the ground-state phase diagram in Figs.~\ref{fig:main}(b) and \ref{fig:main}(c) once additional interactions are included. We therefore next consider the effects of further interactions, namely the $\Gamma'$ interaction and the third-neighbor Heisenberg interaction, which are introduced in Sec.~\ref{sec:modelmethod} and are believed to be subdominant in candidate materials. 
To this end, we carried out PFFRG calculations including these interactions in the frustrated regions.

We first discuss the effect of the $\Gamma'$ interaction. The resulting susceptibility profiles are shown in Fig.~\ref{fig:Gprime}. As shown in Figs.~\ref{fig:Gprime}(a) and \ref{fig:Gprime}(c), for $\Gamma' > 0$, both parameter sets exhibit N\'eel ordered states. In contrast, for $\Gamma' < 0$, Fig.~\ref{fig:Gprime}(b) exhibits the $\mathrm{V1}$ state, while Fig.~\ref{fig:Gprime}(d) shows a double-$Q$ zigzag state. The latter corresponds to a magnetic order that can be viewed as a superposition of zigzag orders along two distinct directions, and does not appear in our phase diagrams in the absence of $\Gamma'$ (Fig.~\ref{fig:main}). This state is consistent with the spin structure factor observed in experiments on Na$_2$IrO$_3$~\cite{HwanChun2015} and Na$_3$Co$_2$SbO$_6$~\cite{Gu2024}. Indeed, the parameter set in Fig.~\ref{fig:Gprime}(d) lies close to the magnetic interaction parameters for Na$_2$IrO$_3$ estimated by the quantum chemistry calculations in a previous study~\cite{Katukuri2014}. Furthermore, the importance of a negative $\Gamma'$ for stabilizing zigzag order in the regime where the FM Kitaev interaction is dominant has been pointed out in a previous ED study~\cite{Rau2014a}, and our results are also consistent with this picture.

We next turn to the effect of the third-neighbor Heisenberg interaction, $J_3$, in the highly frustrated regions. Figure~\ref{fig:J3} shows the PFFRG results for the model including the additional term in Eq.~\eqref{eq:J3}. 
Figures~\ref{fig:J3}(b) and \ref{fig:J3}(c) exhibit the FM and zigzag (more precisely, single-$Q$ zigzag) states, respectively, both of which also appear in our ground-state phase diagram with $J_3=0$ (Fig.~\ref{fig:main}). 
In contrast, Figs.~\ref{fig:J3}(a) and \ref{fig:J3}(d) show triple-$Q$ zigzag and double-$Q$ stripy orders, respectively. The former can be interpreted as a superposition of zigzag orders along three distinct directions, whereas the latter corresponds to a superposition of stripy orders along two distinct directions. Notably, the triple-$Q$ zigzag state is consistent with that suggested by experimental studies on Na$_2$Co$_2$TeO$_6$~\cite{Chen2021, Yao2023, Kruger2023}.

Our results highlight the importance of subdominant interactions in the highly frustrated regions identified in this study. We clarify that these additional interactions give rise to nontrivial multiple-$Q$ ordered states, which are absent in the minimal Kitaev-Heisenberg-$\Gamma$ model, and reflect strong competition among different ordering tendencies. 
In particular, our results demonstrate that incorporating interactions beyond the Kitaev-Heisenberg-$\Gamma$ model, such as $\Gamma'$ and $J_3$, is essential for explaining the magnetic orders observed experimentally.

\section{Summary} \label{sec:summary}
In summary, we have revisited the ground-state phase diagram of the Kitaev-Heisenberg-$\Gamma$ model on the honeycomb lattice from both classical and quantum perspectives by using state-of-the-art numerical techniques.
In the classical limit, employing large-scale gradient descent minimization based on automatic differentiation, we established a comprehensive phase diagram and uncovered a rich variety of multiple-$Q$ magnetically ordered phases beyond the conventional FM, N\'eel, zigzag, stripy, and vortex phases. In particular, we identified a \textit{zoo of noncollinear magnetic orders}, including both commensurate and incommensurate orders, in parameter regions where the Luttinger-Tisza method breaks down. These results demonstrate that the classical Kitaev-Heisenberg-$\Gamma$ model hosts a much richer landscape of competing magnetic states than previously recognized.

In the quantum case, we systematically mapped out the phase diagram by calculations for large system size using the PFFRG method. We found that quantum fluctuations strongly suppress many of the competing magnetic orders present in the classical limit, resulting in a substantially reduced set of dominant magnetic phases. In addition to the phases inherited from the Kitaev-Heisenberg model, we identified several incommensurate magnetically ordered phases, including the newly found $\mathrm{IC3}$ and several intermediate phases. We also revealed highly frustrated regions characterized by suppressed critical cutoff scales and broad momentum-space degeneracy of the spin susceptibility. The susceptibility profiles in these regions exhibit ring-like structures reminiscent of spiral spin liquids, suggesting proximity to unconventional quantum states beyond the magnetic orders stabilized within the present model.

Motivated by these highly frustrated regions, we further investigated the effects of subdominant interactions relevant to candidate materials. We showed that additional interactions such as $\Gamma'$ and the third-neighbor Heisenberg interaction $J_3$ readily lift the near degeneracy and stabilize magnetic orders that are absent from the Kitaev-Heisenberg-$\Gamma$ model, including multiple-$Q$ zigzag and stripy states. These results provide a natural explanation for several experimentally observed magnetic structures and highlight the crucial role of subdominant interactions in understanding the magnetic properties of Kitaev materials.

Overall, our results provide a comprehensive and unified picture of the Kitaev-Heisenberg-$\Gamma$ model across classical and quantum regimes. They provide a valuable guide not only for interpreting experimental observations in existing Kitaev candidate materials but also for the search and design of novel materials hosting Kitaev spin liquids and related frustrated quantum phases. 
Our results also demonstrate that the combination of gradient descent optimization based on automatic differentiation and the PFFRG method is powerful approaches for elucidating phase competition and quantum effects in highly frustrated spin systems.
Since our results underscore the importance of subdominant interactions, a systematic understanding of their roles remains an essential open problem in bridging the gap between minimal spin models and real Kitaev materials.
Another interesting avenue is to explore the impact of quantum fluctuations beyond the one-loop PFFRG approximation employed in the present study. In particular, it will be important to examine whether higher-order contributions, potentially enhanced by additional interactions such as ring-exchange terms, can stabilize more exotic magnetic orders or even other spin-liquid-like states in the highly frustrated regions identified here.

\begin{acknowledgments}
The authors thank K. Ido, Y. Kato, T. Misawa, J. Nasu, and T. Okubo for fruitful discussions. K.F. also thanks M. Gohlke, S. Ikegami, and R. Pohle for constructive suggestions, S. Kasai and K. Shimizu for technical advices, and R. Iwazaki and R. Kaneko for helpful information on relevant literature.
Parts of the numerical calculations have been done using the facilities of the Supercomputer Center, the Institute for Solid State Physics, the University of Tokyo, the Information Technology Center, the University of Tokyo, and the Center for Computational Science, University of Tsukuba.
This work was supported by the Japan Society for the Promotion of Science (JSPS) KAKENHI Grant Nos. JP20H00122, JP24K17009, JP25H01247, and JP26H02011.
\end{acknowledgments}

\appendix
\section{Details of method for classical spins} \label{app:classical}

\begin{figure*}
    \centering
     \includegraphics[width=\linewidth,clip]{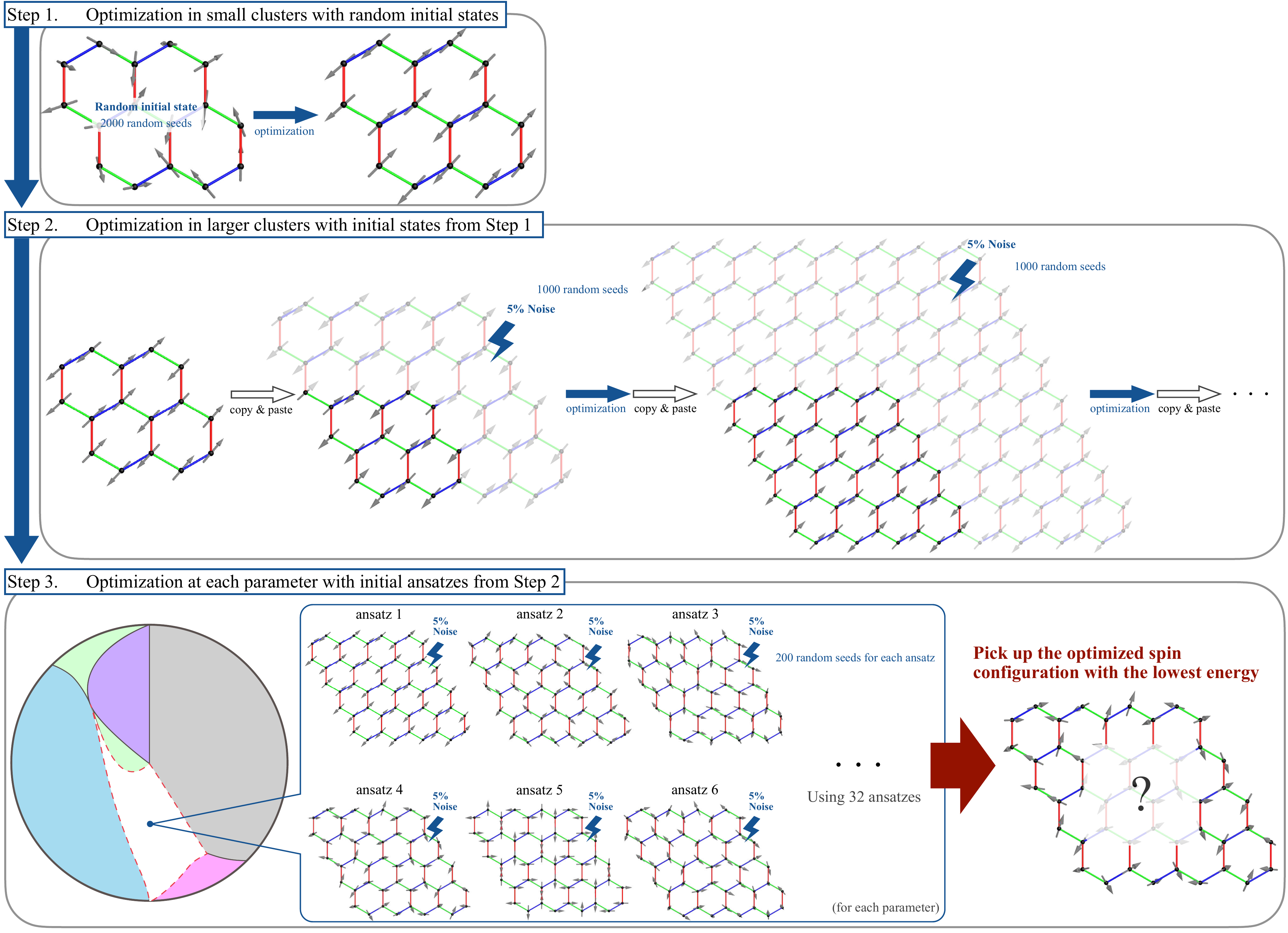}
    \caption{
        Schematic illustration of the three-step optimization procedure used in the classical spin calculations. 
    }
    \label{fig:method}
\end{figure*}

In this section, we describe the details of the optimization procedure used in the classical spin calculations. The overall workflow consists of three steps, as illustrated in Fig.~\ref{fig:method}. Steps 1 and 2 are devoted to the preparation of initial ansatzes, whereas Step 3 involves optimization calculations at each parameter point in the phase diagram using prepared ansatzes in Steps 1 and 2. 

In Step~1, we perform energy minimization on small clusters starting from $2000$ random initial spin configurations generated with different random seeds. For each initial spin configuration, as mentioned in the main text, the optimization is performed using the JAX~\cite{jax} and Optax~\cite{optax}, which provide automatic differentiation and optimization algorithms widely used in machine learning. We employ the Adam optimizer~\cite{adam}, with a total of $N_{\mathrm{step}}=1000$ optimization steps for each run. 
Throughout the optimization runs, the learning rate is scheduled using a cosine-decay scheme, where the learning rate at the $m$-th step is given by 
\begin{align}
    \frac{1}{2}\biggl\{1+\cos\biggl(\frac{\pi m}{N_{\mathrm{step}}}\biggr)\biggr\}.
\end{align}

In Step~2, we progressively increase the system size by copying and tiling the optimized configurations obtained in Step~1 (what is called `copy and paste') to construct initial states for larger clusters. After each tiling, we introduce $5\%$ random noise to the spin angles $(\vartheta_i, \varphi_i)$ to prevent the optimization from becoming trapped in local minima inherited from the smaller systems. We use $1000$ random seeds at each size-increase step. This step is iterated until the target system size $L = 36$ is reached. We perform the above procedure along three system-size sequences, namely $L=2 \rightarrow 6 \rightarrow 12 \rightarrow 36$, $L=3 \rightarrow 6 \rightarrow 12 \rightarrow 36$, and $L=4 \rightarrow 12 \rightarrow 36$. From each sequence, the spin configuration with the lowest energy is selected and incorporated as one of the initial ansatzes. By repeating Step 2 at several tens of parameter points in the phase diagram, we obtain a total of 32 initial ansatzes.

In Step 3, we perform the final optimization at each parameter point $(\theta, \phi)$ in the phase diagram. We employ the 32 initial ansatzes prepared in Steps 1 and 2, and for each ansatz, generate $200$ distinct initial configurations by introducing $5\%$ random noise. 
Optimization is then carried out starting from all $32 \times 200$ initial configurations, and an optimized spin configuration with the lowest energy is selected as the final result at each parameter point. 
This procedure enables a thorough exploration of the energy landscape and suppresses the effects of local minima, particularly in regions of strong competition between different orders.
In Steps 2 and 3, the learning rate is scheduled according to a warmup cosine-decay scheme,
\begin{align}
    \begin{cases}
        \frac{m}{0.1 N_{\mathrm{step}}}  & (0 \leq m \leq 0.1 N_{\mathrm{step}}),\\ \\
        \frac{1}{2}\biggl\{1+\cos\biggl(\frac{m - 0.1 N_{\mathrm{step}}}{0.9 N_{\mathrm{step}}}\pi\biggr)\biggr\} & (0.1 N_{\mathrm{step}} \leq m \leq N_{\mathrm{step}}).
    \end{cases}
\end{align}
 
\section{Eigenvalues of spin inertia tensor} \label{app:inertia}

\begin{figure*}
    \centering
     \includegraphics[width=\linewidth,clip]{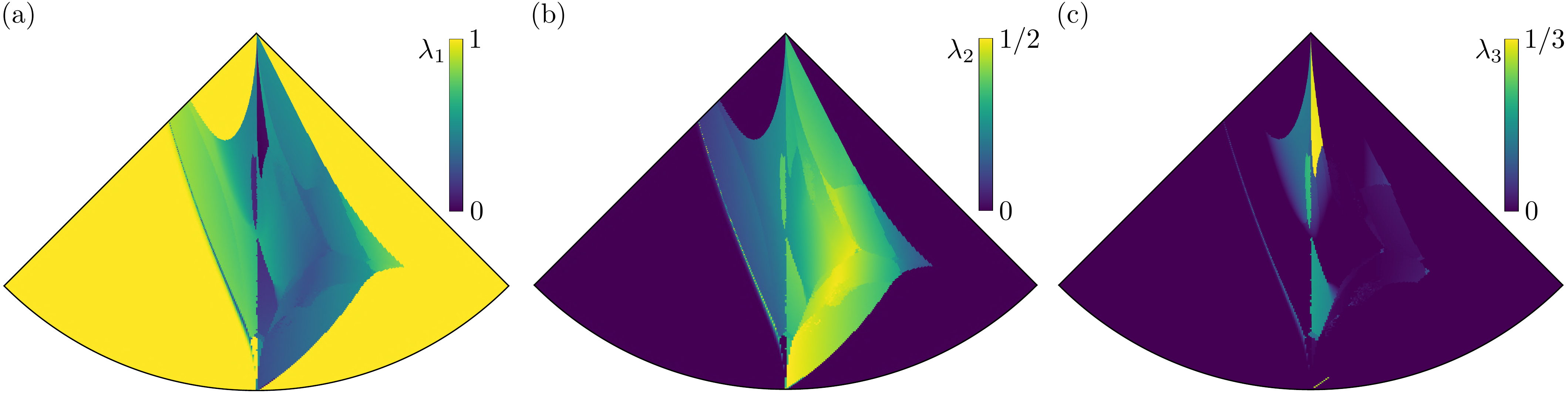}
    \caption{Distribution of the three eigenvalues (a) $\lambda_1$, (b) $\lambda_2$, and (c) $\lambda_3$ of the spin inertia tensor $M^{\alpha\beta}$ over the lower quadrant of the classical phase diagram. The eigenvalues are normalized so that $\lambda_1 + \lambda_2 + \lambda_3 = 1$.
    }
    \label{fig:inertia_tensor}
\end{figure*}

To further support the classification of spin configurations into collinear, coplanar, and noncoplanar orders, we analyze the eigenvalues of the spin inertia tensor~\cite{Sklan2013}. The spin inertia tensor is defined as
\begin{align}
    M^{\alpha\beta} = \frac{1}{N_{\mathrm{site}}} \sum_i S_i^\alpha S_i^\beta,
\end{align}
where $\alpha, \beta \in \{x, y, z\}$. This real symmetric $3\times 3$ matrix has three non-negative eigenvalues $\lambda_1 \geq \lambda_2 \geq \lambda_3 \geq 0$ with $\lambda_1 + \lambda_2 + \lambda_3=1$. These eigenvalues provide a quantitative criterion for distinguishing the nature of the spin configuration: for a collinear state, spins align along a single axis, giving $\lambda_1=1$ and $\lambda_2 = \lambda_3 = 0$; for a coplanar state, spins are confined to a plane, giving $\lambda_3 = 0$ with $\lambda_1, \lambda_2 > 0$; for a noncoplanar state, all three eigenvalues are positive, $\lambda_1, \lambda_2, \lambda_3 > 0$.
 
Figure~\ref{fig:inertia_tensor} shows the distribution of the three eigenvalues over the lower quadrant of the phase diagram. The regions with $\lambda_3 > 0$ are consistent with the noncoplanar phases identified from the scalar spin chirality in Fig.~\ref{fig:classical_data}(b) of the main text, confirming the presence of the noncoplanar ordered phases $18$, $\mathcal{R}(18)$, and $\mathcal{R}(6)$. The eigenvalue analysis also corroborates that the phases $6$, $6'$, $8$, $8'$, $12$, and $24$ are coplanar ($\lambda_3 \simeq 0$, $\lambda_1, \lambda_2 > 0$), while the conventional FM, N\'eel, zigzag, and stripy phases are collinear ($\lambda_2, \lambda_3 \simeq 0$).

\section{Additional PFFRG data} \label{app:pffrg}

\begin{figure*}
    \centering
     \includegraphics[width=\linewidth,clip]{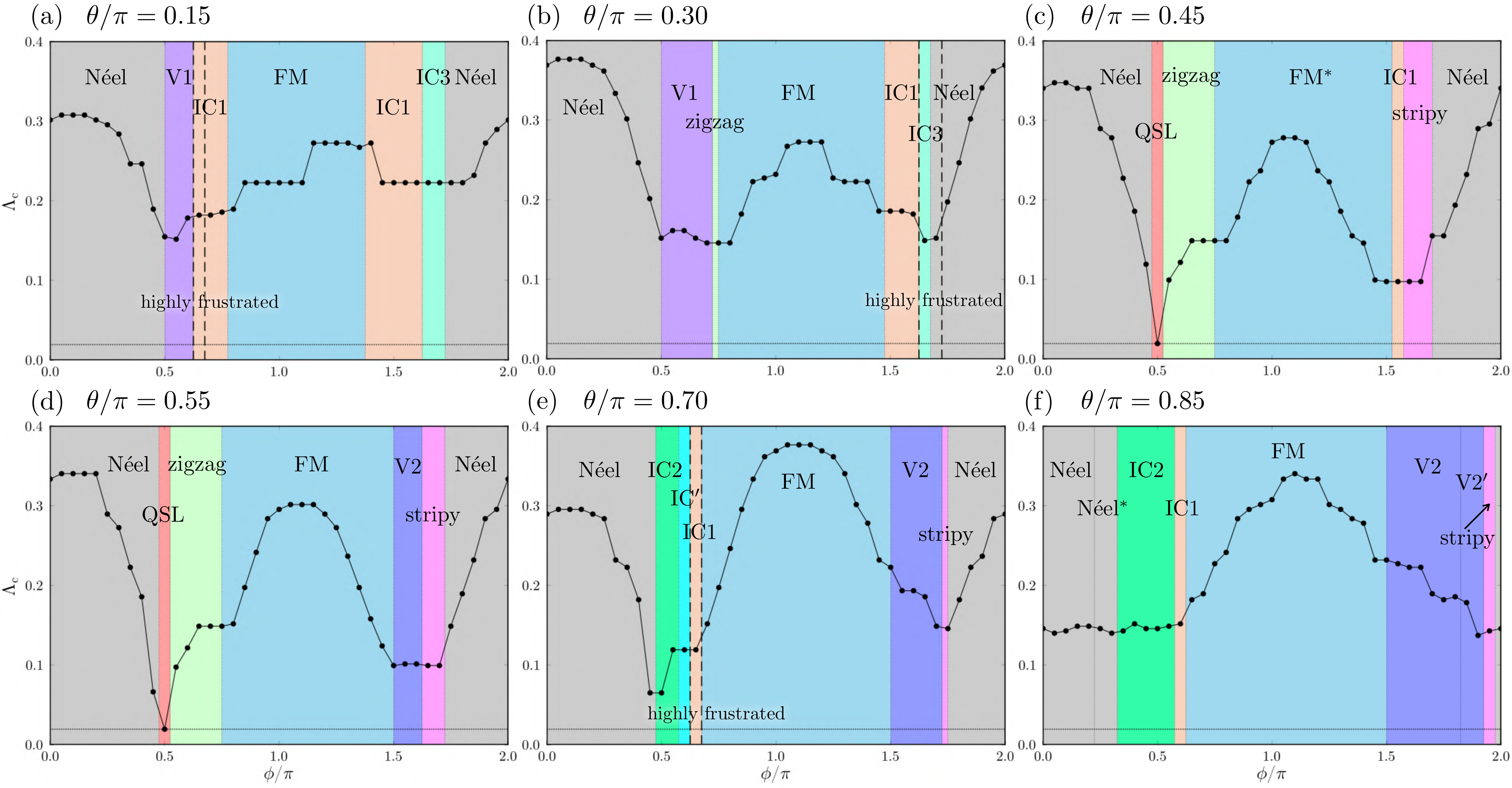}
    \caption{$\phi$ dependences of the critical cutoff energy scale $\Lambda_{\mathrm{c}}$ at fixed values of (a) $\theta/\pi = 0.15$, (b) $0.30$, (c) $0.45$, (d) $0.55$, (e) $0.70$, and (f) $0.85$. The black dots indicate the calculated parameter points. The colored backgrounds denote the phases identified in the main text, following the same color convention as in Figs.~\ref{fig:main}(b) and \ref{fig:main}(c). The vertical black dashed lines correspond to the boundaries of the highly frustrated regions in Figs.~\ref{fig:main}(b) and \ref{fig:main}(c). The dotted horizontal line indicates $\Lambda_{\mathrm{min}} \simeq 10^{-1}$.
    }
    \label{fig:Lambad_cut}
\end{figure*}

Figure~\ref{fig:Lambad_cut} shows the $\phi$ dependence of the critical cutoff energy scale $\Lambda_{\mathrm{c}}$ at several values of $\theta$. These cuts provide a complementary view of the phase diagram in Figs.~\ref{fig:main}(b) and \ref{fig:main}(c) as well as Figs.~\ref{fig:degen_data}(a) and \ref{fig:degen_data}(b) of the main text, allowing a more detailed comparison of the phase boundaries and a confirmation of the suppression of $\Lambda_{\mathrm{c}}$ in the highly frustrated regions.
 
\begin{figure*}
    \centering
     \includegraphics[width=\linewidth,clip]{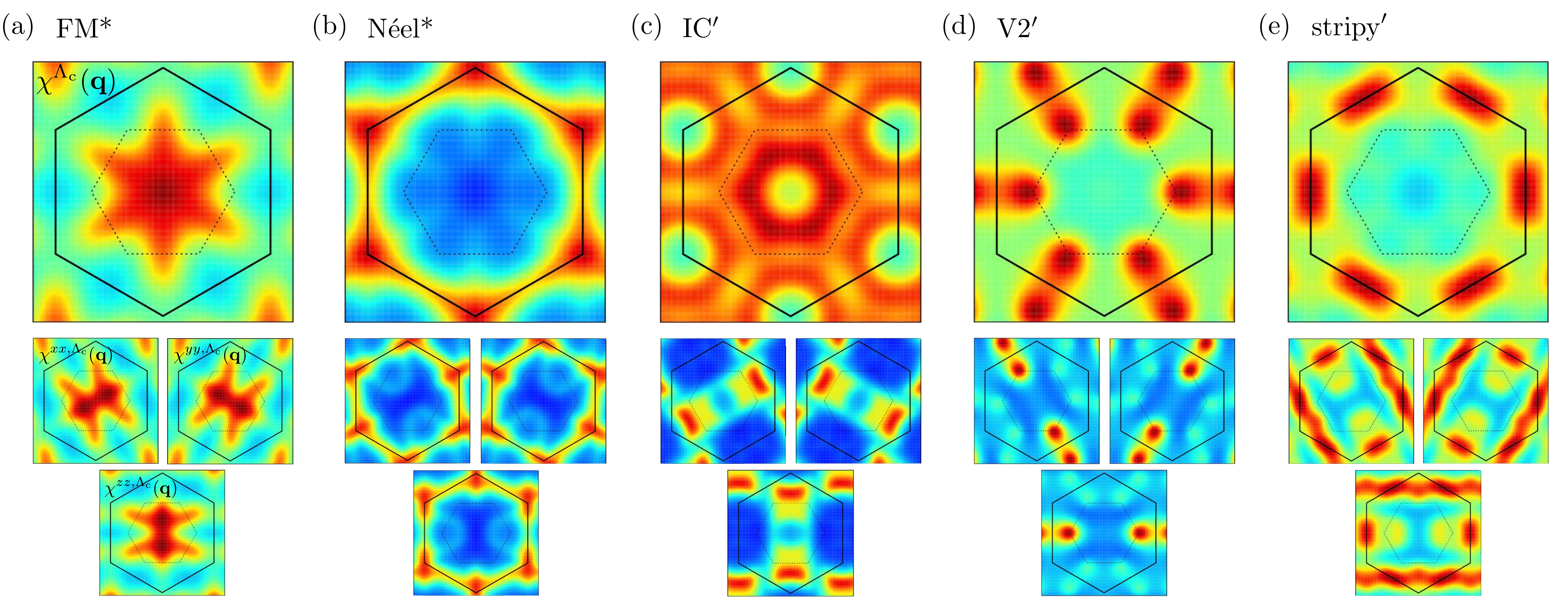}
    \caption{$\mathbf{q}$ dependences of the spin susceptibilities for the subsidiary ordered phases: (a) FM$^*$, (b) N\'eel$^*$, (c) $\mathrm{IC}'$, (d) $\mathrm{V2}'$, and (e) stripy$'$. The left panels show $\chi^{\Lambda_{\mathrm{c}}}(\mathbf{q})$, and the right small panels show $\chi^{xx,\Lambda_{\mathrm{c}}}(\mathbf{q})$, $\chi^{yy,\Lambda_{\mathrm{c}}}(\mathbf{q})$, and $\chi^{zz,\Lambda_{\mathrm{c}}}(\mathbf{q})$. The notations are common to those in Fig.~\ref{fig:pffrg_chiq} in the main text.
    }
    \label{fig:pffrg_add}
\end{figure*}

Figure~\ref{fig:pffrg_add} shows the $\mathbf{q}$ dependences of the spin susceptibility $\chi^{\Lambda_{\mathrm{c}}}(\mathbf{q})$ and its components $\chi^{xx,\Lambda_{\mathrm{c}}}(\mathbf{q})$, $\chi^{yy,\Lambda_{\mathrm{c}}}(\mathbf{q})$, and $\chi^{zz,\Lambda_{\mathrm{c}}}(\mathbf{q})$ for the subsidiary ordered phases discussed in Sec.~\ref{subsec:result_quantum} of the main text. Figures~\ref{fig:pffrg_add}(a) and \ref{fig:pffrg_add}(b) show the results for the FM$^*$ and N\'eel$^*$ phases, respectively. In these phases, the total susceptibility $\chi^{\Lambda_{\mathrm{c}}}(\mathbf{q})$ exhibits the same peak positions as in the conventional FM and N\'eel phases, while the component-resolved susceptibilities develop peaks at shifted wave vectors. Figure~\ref{fig:pffrg_add}(c) shows the $\mathrm{IC}'$ phase, which appears as an intermediate phase between $\mathrm{IC1}$ and $\mathrm{IC2}$ with distinct peak positions in $\chi^{\Lambda_{\mathrm{c}}}(\mathbf{q})$. 
Figure~\ref{fig:pffrg_add}(d) shows the $\mathrm{V2}'$ phase, where the susceptibility peaks are slightly shifted from those of the $\mathrm{V2}$ phase toward incommensurate wave vectors. Figure~\ref{fig:pffrg_add}(e) shows the stripy$'$ phase, which shares similar peak positions in $\chi^{\Lambda_{\mathrm{c}}}(\mathbf{q})$ with the stripy phase but exhibits dominant peaks along different edges of the Brillouin zone in $\chi^{\gamma\gamma,\Lambda_{\mathrm{c}}}(\mathbf{q})$.

\bibliography{library}

\end{document}